\documentclass[11pt,a4paper]{article}

\usepackage{amsmath, amsfonts, amssymb}

\usepackage[T1]{fontenc}
\usepackage[utf8]{inputenc}
\usepackage[english]{babel}
\usepackage{fullpage}

\usepackage{hyperref}

\usepackage[nosort]{cite}
\usepackage{bbm}

\numberwithin{equation}{section}

\DeclareMathOperator{\re}{Re}

\newcommand{\bbR}{{\mathbb{R}}}
\newcommand{\bbC}{{\mathbb{C}}}
\newcommand{\bbZ}{{\mathbb{Z}}}

\newcommand{\bbone}{{\mathbbmss{1}}}

\begin{document}

\begin{titlepage}
	\thispagestyle{empty}
	\begin{flushright}
		\hfill{DFPD/2014-TH/06}\\
		\hfill{Nikhef-2014-010}
	\end{flushright}
	
	\vspace{35pt}
	
	\begin{center}
	    { \huge{\bf\boldmath Symplectic Deformations\\[1ex] of Gauged Maximal Supergravity }}
		
		\vspace{40pt}
		
		{ {{{\bf Gianguido~Dall'Agata$^{1,2}$, Gianluca~Inverso$^3$,}\\and {\bf Alessio Marrani$^{4}$}}}}
		
		\vspace{35pt}
		
		{
		$^1${\it  Dipartimento di Fisica e Astronomia ``Galileo Galilei''\\
		Universit\`a di Padova, Via Marzolo 8, 35131 Padova, Italy}
		
		\vspace{2ex}
		
	    $^2${\it   INFN, Sezione di Padova \\
		Via Marzolo 8, 35131 Padova, Italy}

		\vspace{2ex}
		
		$^3${\it Nikhef, Science Park 105, 1098 XG Amsterdam, The Netherlands}
		}		

       \vspace{2ex}
		
		$^4${\it Instituut voor Theoretische Fysica,\\ KU Leuven,
Celestijnenlaan 200D, B-3001 Leuven, Belgium}

		\vspace{15ex}
		
		{ABSTRACT}
	\end{center}
	
We identify the space of symplectic deformations of maximal gauged supergravity theories.
Coordinates of such space parametrize inequivalent supergravity models with the same gauge group.
We apply our procedure to the SO(8) gauging, extending recent analyses.
We also study other interesting cases, including Cremmer--Scherk--Schwarz models and gaugings of groups contained in SL$(8,{\mathbb R})$ and in SU$^*(8)$.
	
	\vspace{10pt}

\end{titlepage}

\tableofcontents

\section{\label{Intro}Introduction and summary of main results}

Supergravity theories play a prominent role in revealing many features of string theory, addressing physics beyond the Standard Model and understanding ultraviolet properties of perturbative quantum gravity. 
Explaining the structure of supergravity models and their relation with effective theories of strings is therefore a task of primary importance, which is unfortunately quite far from completion.
We are closer to achieving this goal for maximal supergravities, because maximal supergravities have a unique multiplet, very constrained couplings, and gauge interactions are the only known way to generate masses and a scalar potential.

A very interesting aspect of the gauging procedure is the existence of an infinite number of consistent models with different couplings for given gauge groups \cite{Dall'Agata:2012bb}.
This recent discovery makes even more compelling a thorough review of the structure of maximal gauged supergravities, especially in view of their stringy origin and of the interpretation of their anti--de Sitter vacua in terms of the gauge/gravity duality.
For instance, it is well-known that the original ${\rm SO(8)}$ gauged maximal supergravity \cite{deWit:1981eq,deWit:1982ig} can be regarded as the consistent truncation of eleven-dimensional supergravity compactified on a seven-sphere \cite{de Wit:1986iy} (see also  \cite{Nicolai:2011cy,deWit:2013ija,Godazgar:2013pfa} for recent developments of the original analysis), which in turn is dual to the ABJM theory (for Chern--Simons level $k=1$) \cite{Aharony:2008ug} in the large $N$ limit.
However, we now know that there is a continuous deformation parameter (often denoted as $\omega$), which changes the couplings of this model, preserving the maximally supersymmetric AdS vacuum \cite{Dall'Agata:2012bb}. 
If, on the one hand, it is difficult to imagine an infinite number of string backgrounds with SO(8) symmetry, on the other hand, it is even more challenging to understand the meaning of such deformations in the ABJM theory.

Since \cite{Dall'Agata:2012bb}, many different analyses of the properties of the `$\omega$-deformed' ${\rm SO(8)}$ gauged supergravities have been carried out through the study of several further truncations, studying in particular maximally symmetric vacua, domain walls and black hole solutions \cite{Borghese:2012qm,Borghese:2012zs,Borghese:2013dja,Guarino:2013gsa,Tarrio:2013qga,Anabalon:2013eaa,Lu:2014fpa}.
At the same time, analogous $\omega$-deformations for non-compact ${\rm SO}(p,q)$ gaugings have been identified and used to show that it is possible to embed slow-roll scenarios in gauged maximal supergravity \cite{Dall'Agata:2012sx}, and to study the moduli space of Minkowski models of maximal supergravity with spontaneously broken supersymmetry \cite{Dall'Agata:2012cp,Catino:2013ppa}.
The fact that similar deformations exist for several gaugings and that the $\omega$ parameter often survives the truncation to models with lower supersymmetry suggests that such deformations of gauged supergravity can be a quite general phenomenon, and not limited to the maximal theory.

Physically, $\omega$ corresponds to the possibility of deforming the couplings of a gauged supergravity action by changing the symplectic embedding of the vector fields of the theory that give rise to the gauge connection, in a way that preserves compatibility with the structure of the gauge group.
This deformation of the symplectic embedding affects the couplings with other fields, as well as the supersymmetry variations and the scalar potential.

Given the diversity of physical interpretations and effects that these deformations can have, it is important to understand how they can be rigorously defined and classified.
This type of analysis would also play a crucial role in any attempt to classify all the allowed gaugings of a supergravity theory.
Moreover, a consistent definition of such deformations should make it possible to clearly identify the correct range of inequivalence of the deformation parameter(s), which is an important point on which there has been some confusion in the literature.

In this paper we focus on maximal supergravity in $D=4$ and we describe how to characterize these deformations in full generality.
We will define the appropriate space of `symplectic deformations' in terms of the allowed (local and non-local) field redefinitions and dualities of the maximal theory, using the embedding tensor formalism \cite{Nicolai:2000sc,deWit:2007mt} in order to perform a general analysis that can be applied to any gauging.

Let us give a brief preview of our general results.
For ungauged maximal supergravity, the set of Lagrangians that cannot be mapped to each other by local field redefinitions is identified with the double quotient space \cite{deWit:2002vz}
\begin{equation}
\label{inequivalent Lagrangians}
\rm GL(28,\bbR)\ \backslash\ Sp(56,\bbR)\ /\ E_{7(7)}.
\end{equation}
Local field redefinitions of the 28 vector fields of the theory correspond to the $\rm GL(28,\bbR)$ quotient.
The (continuous version of the) U-duality group of maximal supergravity in $D=4$ is $\rm E_{7(7)}$, which also corresponds to the isometry group of the scalar manifold $\rm E_{7(7)}/SU(8)$.
In fact, what appears in the right quotient of \eqref{inequivalent Lagrangians} must be regarded as local redefinitions of the scalar fields by these isometries, as opposed to $\rm E_{7(7)}$ dualities which must also act on vector fields. 
The different Lagrangians correspond to distinct `symplectic frames' and are invariant under different `electric' subgroups of the $\rm E_{7(7)}$ duality group acting locally on the physical fields.
The resulting equations of motion and Bianchi identities are equivalent for any Lagrangian defined by \eqref{inequivalent Lagrangians}.

When we turn on a gauging, the quotient \eqref{inequivalent Lagrangians} still parameterizes a set of consistent Lagrangians, provided that we let $\rm Sp(56,\bbR)$ also act
on the `gauging parameters', defined in terms of the embedding tensor formalism as a set of generators $(X_M)_N{}^P \in {\mathfrak e_{7(7)}}$.
The resulting theories are again equivalent at the level of the equations of motion.
There is instead a set of symplectic transformations that can act on the couplings of the theory as in the ungauged case, not acting on $X_M$, and still give a fully consistent gauged supergravity.
We dub these transformations `symplectic deformations', and we will prove that they provide the correct generalization of the $\omega$-deformation of the $\rm SO(8)$ theory.
The space of symplectic deformations is the normalizer of the gauge group $\mathcal N_{\rm Sp(56,\bbR)}(\rm G_{gauge})$, quotiented by a proper set of transformations that can be reabsorbed in field redefinitions.
If we define our gauged theory in an electric frame, or alternatively if we integrate out and gauge fix the extra vector and tensor fields that may appear in a generic choice of symplectic frame, effectively switching back to an electric frame \cite{deWit:2005ub}, then we have a consistent notion of local redefinitions of the physical vector fields, and we can quotient by them together with redefinitions of the scalars.
The space of inequivalent deformations turns out to be
\begin{equation}
\label{S space INTRO}
\mathfrak S
 \ \equiv\
 {\cal S}_{{\rm GL(28,\bbR)}}(X)
 \ \backslash\
 {\cal N}_{{\rm Sp(56,\bbR)}}({\rm G_{gauge}})\ /\
 {\cal N}_{\bbZ_2\ltimes{\rm E_{7(7)}}}({\rm G_{gauge}}),
\end{equation}
where ${\cal N}_{\rm G}({\rm G_{gauge}})$ is the normalizer of ${\rm G_{gauge}}$ in G, while ${\cal S}_{{\rm GL(28,\bbR)}}(X)$ is the group of ${\rm GL(28,\bbR)}$ transformations that stabilize $X_{MN}{}^P$ up to overall rescalings.
With this definition we do not discriminate between theories that differ only in the value of the gauge coupling constant.
However, if we insist on regarding them as distinct models, we can simply take the left denominator in \eqref{S space INTRO} to be the stabilizer of $X_{MN}{}^P$ in $\rm GL(28,\bbR)$.
The $\bbZ_2$ factor in the right quotient denotes the outer automorphism of $\rm E_{7(7)}$, whose action is strictly related to a parity transformation \cite{Ferrara:2013zga,TrigianteTOAPPEAR}, and is quite subtle in this context.
The precise definitions will be given in the next sections.
In some cases, including G$_{\rm{gauge}} = {\rm SO(8)}$ in the standard $\rm SL(8,\bbR)$ frame, we find that the classification of symplectic deformations can be carried out using group theoretical methods exclusively.
In this way, we will re-analyze the SO(8) case in detail as an instructive exercise, also providing a complementary proof that the range of $\omega$ is $[0,\pi/8]$ \cite{Dall'Agata:2012bb,deWit:2013ija}.

We stress that the definition of $\mathfrak S$ depends on the choice of electric frame, because the set of local field redefinitions depends on this choice.
However, some of the transformations in $\mathfrak S$ do not affect the symplectic embedding of the gauge connection and as a consequence they  do not affect the equations of motion.
For example, we find that \eqref{S space INTRO} consistently encodes the fact that the standard electric action of $\rm SO(8)$ gauged maximal supergravity admits the introduction of a field-independent, gauge invariant shift in the $\theta$-angle of the (gauged) field strengths.
Even if such terms can be physically relevant at the quantum level, we can choose to define a `reduced' $\mathfrak S$-space that is completely independent from the choice of symplectic frame and classifies all and only the deformations that do affect the equations of motion.
This space is
\begin{equation}
\label{reduced S space INTRO}
\mathfrak S_{\rm red}\ \equiv\ {\cal S}_{{\rm Sp(56,\bbR)}}(X)\ \backslash\  {\cal N}_{{\rm Sp(56,\bbR)}}({\rm G_{gauge}})\ /\ {\cal N}_{\bbZ_2\ltimes{\rm E_{7(7)}}}({\rm G_{gauge}}).
\end{equation}
This definition treats as equivalent also those Lagrangians that are mapped to each other by changes of symplectic frame that do not affect the gauge connection nor any coupling induced by the gauging (namely, they stabilize $X$).
If we go back to the ungauged theory setting $X=0$, $\mathfrak S$ matches \eqref{inequivalent Lagrangians} while $\mathfrak S_{\rm red}$ becomes trivial.

We will combine these tools with a convenient adaptation of the embedding tensor formalism, which allows us to reduce the problem of identifying all consistent gauge connections for a given gauge group to a set of linear equations.
These techniques will then make it easy to identify several new examples of symplectic deformations.
We will analyze all gauge groups contained in $\rm SL(8,\bbR)$ and $\rm SU^\ast(8)$, as well as the Cremmer--Scherk--Schwarz (CSS) gaugings.
For the latter no `$\omega$-deformation' turns out to be possible, while new examples of deformations, with interesting physical effects, are found for the gaugings of ${\rm ISO}(p,7-p)$ and of real forms of ${\rm SO(4,\bbC)}^2\ltimes T^{16}$.
We will also identify the resulting ranges for the deformation parameters.

\section{\label{Consistency}Consistency constraints on gauge connection}

The gauging process promotes up to 28 of the vector fields $A_\mu^M$, transforming in the $\mathbf{56}$ representation of the E$_{7(7)}$ duality group, to connection fields for the gauge group $\mathrm{G_{gauge}}$.
Consistency of the procedure requires that the corresponding generators $X_M$ satisfy the constraints \cite{deWit:2007mt}
\begin{equation}  \label{X constraints}
[X_M,\,X_N]=-X_{MN}{}^P\ X_P,\qquad X_{MN}{}^M = X_{(MNP)}=0,
\end{equation}
where $(X_M)_N{}^P = X_{MN}{}^P$ are the gauge generators in the $\mathbf{56}$ representation and $X_{MNP}=X_{MN}{}^Q \Omega_{PQ}$. 
The embedding tensor formalism relates the gauge generators to the elements of the ${\mathfrak e}_{7(7)}$ algebra by introducing the $\Theta$ tensor: $X_M = \Theta_M{}^\alpha t_\alpha$.
One can therefore translate the consistency conditions (\ref{X constraints}) in terms of constraints on $\Theta$.
However, once we fix a choice of gauge algebra we can also introduce $\mathrm{G_{gauge}}$ adjoint indices $r,s,\ldots$, so that the gauge generators are $t_r$, and write
\begin{equation}  \label{vth definition}
X_M = \vartheta_M{}^r t_r,\quad r,s,\ldots = 1, \ldots, \mathrm{dim}(\mathrm{%
G_{gauge}})\leq28.
\end{equation}
The constraints now read:
\begin{align}
&[t_r, t_s]=f_{rs}{}^t t_t, & &f_{[rs}{}^v f_{t]v}{}^u=0, \\
&\vartheta_M{}^s f_{rs}{}^t = - t_{r\,M}{}^N \vartheta_N{}^t, & &
\vartheta_{M}{}^r t_{rN}{}^M = \vartheta_{(M}^r t^{\vphantom{r}}_{rNP)}=0.
\label{vth constraints}
\end{align}
Given a Lie subalgebra $\{t_r\}\subset \mathfrak{e}_{7(7)}$ of dimension $\dim\mathrm{G_{gauge}}\le28$, any solution of \eqref{vth constraints} provides a consistent gauging. 
The above constraints are exhaustive and guarantee the consistency of the gauging. 
In particular, after we solve these constraints, locality is guaranteed to hold, i.e.\begin{equation}
\vartheta_M{}^r\vartheta_N{}^s\Omega^{MN}=0.
\end{equation}

Given a consistent gauging defined by some $X^0_{MN}{}^P$, we can always choose an initial symplectic frame such that $X^0$ is electric, and by a choice of basis of the gauge generators we can then set without loss of generality
\begin{equation}  \label{X0}
X^0_{MN}{}^P = \delta_M{}^r t_{r\,N}{}^P,\quad r=1,\ldots\,,\dim\mathrm{%
G_{gauge}}.
\end{equation}
When $\dim\mathrm{G_{gauge}}<28$ it is useful to introduce indices $a,b,\ldots $ running among the (electric) vector fields $A^a_\mu$ that do not take part in the gauge connection defined by \eqref{X0}. 
Then $t_r$ take the general form \cite{deWit:2007mt}:
\begin{equation}  \label{electric gauge generators}
t_{r\,M}{}^{N} =
\begin{pmatrix}
-f_{rs}{}^t & h_{rs}{}^a & C_{rst} & C_{rsa} \\
0 & 0 & C_{rtb} & 0 \\
0 & 0 & f_{rt}{}^s & 0 \\
0 & 0 & -h_{rt}{}^b & 0 
\end{pmatrix}%
,
\end{equation}
where $f_{rs}{}^t$ are the structure constants of the gauge algebra and $C_{(rst)} = C_{r[st]} = C_{(rs)a} = h_{(rs)}{}^a = 0$. 
The constraints in (\ref{vth constraints}) now become
\begin{align}
	&\vartheta_r{}^u f_{su}{}^t  - f_{sr}{}^u \vartheta_u{}^t + h_{sr}{}^a \vartheta_a{}^t + C_{sru} \vartheta^{ut} + C_{sra} \vartheta^{at} = 0,
\\[1ex]&
	\vartheta_a{}^u f_{su}{}^t = \vartheta^{ru} f_{su}{}^t = \vartheta^{au} f_{su}{}^t = 0,
\\[1ex]&
	\vartheta_r{}^u f_{su}{}^r + \vartheta_a{}^u h_{us}{}^a + \vartheta^{ru} C_{usr} + \vartheta^{au} C_{usa} = 0,
\\[1ex]&
	\vartheta^{ru} C_{uar} = \vartheta^{su} h_{us}{}^a = \vartheta^{su} f_{us}{}^r = 0,
\\[1ex]&
	\vartheta_{(r}{}^u C_{u st)} =2 \vartheta_{(r}{}^u C_{u s)a} + \vartheta_a{}^u C_{urs} = \vartheta_a{}^u C_{urb} = \vartheta^{tu}C_{urs} = \vartheta^{au} C_{urs} =0.
\end{align}
This form of the generators guarantees that $\vartheta_M{}^r = \delta_M{}^r$ is a solution of the constraints. 

\section{\label{Sympl-Def}Symplectic deformations}

Even when we fix the choice of a gauge group, and hence of $t_r$, there is still the possibility that \eqref{vth constraints} admit more than one solution, leading to gauged supergravities that are potentially inequivalent even if they share the same set of gauge symmetry generators, because they differ in the choice of the (electric and magnetic) vector fields that form the gauge connection.
Our aim is to characterize group-theoretically the space of these inequivalent theories, showing the relation between the set of consistent choices of gauge connections (for fixed $t_r$) and symplectic transformations. 

\subsection{Symplectic maps between gauge connections}

First, we will prove that finding all non-vanishing solutions of \eqref{vth constraints} for a fixed choice of $\mathrm{G_{gauge}}$ is equivalent to identifying $\mathcal{N}_{\mathrm{Sp(56,{\mathbb{R}})}}(\mathrm{G_{gauge}})$, up to its subgroup of transformations that leave $X^0$ invariant.
We are going to show that whenever we find solutions $\vartheta_M{}^r$ to \eqref{vth constraints} for the same set of generators $t_r$ (other than $\delta_M{}^r$), this fact can be reinterpreted as the existence of a non-trivial normalizer of the gauge group in $\mathrm{Sp(56,{\mathbb{R}})}$.
Then, the new gauge connections $\vartheta_M{}^r$ define new gauge couplings
$X_{MN}{}^P\equiv \vartheta_M{}^r\,t_{r\,N}{}^P $, and for each $\vartheta_M{}^r$ there exists some element $N\in\mathcal{N}_{\mathrm{Sp(56,{\mathbb{R}})}}(\mathrm{G_{gauge}})$ such that
\begin{equation}  \label{action of normalizer on X} 
	X_{MN}{}^P\ =\ N_M{}^Q\,N_N{}^R\ X^0_{QR}{}^S\ (N^{-1})_S{}^P\,, \qquad N\in\mathcal{N}_{\mathrm{Sp(56,{\mathbb{R}})}}(\mathrm{G_{gauge}}).
\end{equation}

To prove this claim, we start by showing that any element of $\mathcal{N}_{\mathrm{Sp(56,{\mathbb{R}})}}(\mathrm{G_{gauge}})$ defines a consistent connection. 
This is true because the general action of these transformations on $t_r$ reads:
\begin{equation}
\label{normalizer gives GL action on tr}
N_M{}^N\ t_{r\,N}{}^P\ (N^{-1})_P{}^Q = g_r{}^s t_{s\,M}{}^Q,
\quad N\in{\cal N}_{{\rm Sp(56,\bbR)}}({\rm G_{gauge}}),
\quad g\in{\rm GL(\dim{\rm G_{gauge}},\bbR)}.
\end{equation}
We can then define a $\mathrm{GL(28,{\mathbb{R}})}$ transformation
\begin{equation}  \label{GL28 associated with normalizer}
H_M{}^N\equiv%
\begin{pmatrix}
g &  &  &  \\
& q &  &  \\
&  & g^{-T} &  \\
&  &  & q^{-T}%
\end{pmatrix}%
\end{equation}
for some invertible matrix $q$ that does not play any role in the following.
Now, the action of $N$ on the original gauge couplings reads:
\begin{equation}
N_M{}^Q\,N_N{}^R\ X^0_{QR}{}^S\ (N^{-1})_S{}^P = N_M{}^Q H_Q{}^R X^0_{RN}{}^P\,,
\end{equation}
and since we never dropped symplectic covariance, we conclude that the new gauge connection
\begin{equation}
\label{normalizer gives connection}
\vartheta_M{}^r \equiv N_M{}^N H_N{}^P \delta_P{}^r
\end{equation}
satisfies all consistency conditions \eqref{vth constraints}.
Of course, we can set $H=\bbone$ for elements of the centralizer.

Now we must prove that for any solution $\vartheta$ of \eqref{vth constraints} there is some $N_M{}^N$ that yields $\vartheta$ through (\ref{normalizer gives GL action on tr}--\ref{normalizer gives connection}).
First of all, let us define a symplectic matrix $B_M{}^N$ that maps the original gauge connection $\delta_M{}^r$ to some other solution $\vartheta_M{}^r$ of \eqref{vth constraints}:
\begin{equation}  \label{B gives vth}
B_M{}^N \delta_M{}^r = \vartheta_M{}^r.
\end{equation}
Assuming for definiteness that $\dim\mathrm{G_{gauge}}=28$, we can parameterize the most general symplectic $B$ as follows:\footnote{The matrices $\bar\vartheta + (\Omega\vartheta x)^T$, for generic $x$,
parameterize all possible pseudoinverses of $\vartheta$.}
\begin{equation}  \label{B parameterization}
B_M{}^N = \left( \vartheta_M{}^r,\ -(\Omega\bar\vartheta^T)_{M\,r} +
(\vartheta x)_{M\,r}\right), \qquad x_{[rs]} = 0,
\end{equation}
where $\bar\vartheta$ is the (unique) pseudoinverse of $\vartheta$ satisfying
\begin{equation}
\bar\vartheta_r{}^M\ \vartheta_M{}^s = \delta_r{}^s,\quad
\vartheta_M{}^r\bar\vartheta_r{}^N\equiv\pi_M{}^N,\quad \pi_M{}^N=\pi_N{}^M
,\quad \pi_M{}^N\pi_N{}^P = \pi_M{}^P.
\end{equation}
The projector $\pi_M{}^N$ projects orthogonally onto the vector fields of the gauge connection defined by $\vartheta$. 
Equivalently, we can factorize $B$ according to\footnote{In this notation symplectic transformations act on field strengths from the right: $F^M_{\mu\nu}\to F_{\mu\nu}^N B_N{}^M$.}
\begin{equation}  \label{factorize x out of B}
B_M{}^N = \left( \vartheta_M{}^s,\ -(\Omega\bar\vartheta^T)_{M,\,s}\right)
\cdot%
\begin{pmatrix}
\delta^r_s & x_{rs} \\
& \delta^s_r%
\end{pmatrix}%
.
\end{equation}
This construction generalizes to $\dim\mathrm{G_{gauge}}<28$, where there is even more freedom to define a symplectic $B_M{}^N$ satisfying \eqref{B gives vth}.

The `closure' constraint on $\vartheta$ translates into the following property for $B_M{}^N$:
\begin{equation}  \label{almost centralizer}
(B^{-1} X^0_M B)^{\vphantom{0}}_N{}^P \ X_P^0 = X^0_{MN}{}^P \ X_P^0,
\end{equation}
We can see that $B$ `{almost}' centralizes the gauge generators, i.e.~it is only guaranteed that they are centralized by $B$ when further contracted with the (old) embedding tensor. 
When there is one choice of $B_M{}^N$ satisfying \eqref{B gives vth} that actually centralizes $\mathrm{G_{gauge}}$, then clearly the connection to $\mathcal{N}_{\mathrm{Sp(56,{\mathbb{R}})}}(\mathrm{G_{gauge}})$ is proven for this specific case. 
All that is left to
complete our proof is to show that when no choice of $B_M{}^N$ centralizes $\mathrm{G_{gauge}}$, we can nevertheless find an alternative transformation $N_M{}^N$ that normalizes $t_r$ and yields $\vartheta_M{}^r$ through (\ref{normalizer gives GL action on tr}\,--\ref{normalizer gives connection}),
hence proving that the identification of all solutions of \eqref{vth constraints} is equivalent to finding all the elements of $\mathcal{N}_{\mathrm{Sp(56,{\mathbb{R}})}}(\mathrm{G_{gauge}})$ that do not stabilize $ X^0 $.

First of all, notice that the decomposition $X_{MN}{}^P = \vartheta_M{}^r t_{r\,N}{}^P$ is actually redundant, as the only object that really counts in defining a gauging is $X_{MN}{}^P$, or equivalently $X_{MNP}\equiv X_{MN}{}^Q\Omega_{PQ}$. 
The action of $B_M{}^N$ on $X^0_{MNP}$ maps it to another consistent $X_{MNP}$, by virtue of the consistency conditions satisfied by $\vartheta_M{}^r$, which in turn implicitly defines $B_M{}^N$ through \eqref{B gives vth}. 
In particular, the symmetry properties of $X^0_{MNP}$
\begin{equation}  \label{symmetry properties of X0}
X^0_{M[NP]}=0,\qquad X^0_{(MNP)}=0
\end{equation}
are preserved by the action of the linear map $\mathcal{B}_{MNP}{}^{QRS} \equiv B_M{}^Q \delta_N{}^R\delta_P{}^S$, defined as acting on generic three-tensors $T_{MNP}$. 
However, it is clear that the action of $\mathcal{B} $ on tensors orthogonal to $X^0_{MNP}$ will in general not preserve their symmetry properties. 
Since we are only interested in how $\mathcal{B}$ acts on $X^0_{MNP}$, we can always construct a different matrix $\mathcal{N} _{MNP}{}^{QRS}$ by modifying the other entries of $\mathcal{B}$, so that $ \mathcal{N}$ has the same action on $X^0_{MNP}$, but also acts on all other tensors preserving their symmetry properties. 
Therefore $\mathcal{N} _{MNP}{}^{QRS}$ can be factorized and we have
\begin{equation}  \label{from B to N}
\mathcal{N}_{MNP}{}^{QRS} X^0_{QRS} = X_{MNP} = B_M{}^Q X^0_{QNP}, \qquad
\mathcal{N}_{MNP}{}^{QRS} = N_M{}^Q N_N{}^R N_P{}^S.
\end{equation}
Compatibility with the symplectic structure $\Omega_{MN}$ then guarantees that we can choose $N_M{}^N$ to be symplectic, and since $X^0$ and $X$ differ by the choice of gauge connection, but share the same set of generators, we conclude that $N_M{}^N\in\mathcal{N}_{\mathrm{Sp(56,{\mathbb{R}})}}(\mathrm{G_{gauge}})$ and that we could have decomposed $B_M{}^N = N_M{}^P\,H_P{}^N$ from the beginning for at least one symplectic solution of \eqref{B gives vth}, with $H_M{}^N$ defined in (\ref{normalizer gives GL action on tr},\,\ref{GL28 associated with normalizer}). 
This concludes our proof that any non-vanishing solution of \eqref{vth constraints} is associated with an element of $\mathcal{N}_{\mathrm{Sp(56,{\mathbb{R}})}}(\mathrm{G_{gauge}})$ or, more precisely, that their classification (up to overall rescalings) is equivalent to calculating the quotient 
\begin{equation}  \label{normalizer over stabilizer}
\mathcal{S}_{\mathrm{Sp(56,{\mathbb{R}})}}(X^0)\ \backslash\ \mathcal{N}_{%
\mathrm{Sp(56,{\mathbb{R}})}}(\mathrm{G_{gauge}}).
\end{equation}

\subsection{Transformation properties of the Lagrangian}

In order to asses to what extent the choice of different gauge connections classified by \eqref{normalizer over stabilizer} can affect the physics, it is necessary to understand how the couplings of the Lagrangian change with different choices of $\vartheta_M{}^r$.
Since any such choice of gauge connection is associated with an element of $\mathcal{N}_{\mathrm{Sp(56,{\mathbb{R}})}}(\mathrm{G_{gauge}})$, we can always perform a change of symplectic frame in order to map the gauge connection back to its standard electric form, at the price of modifying several other couplings.
Switching to an electric frame associated to each solution of \eqref{vth constraints} can be also reinterpreted as gauge fixing and integrating out the extra vector fields and the two-forms that are included for magnetic gaugings \cite{deWit:2005ub}, so that the Lagrangian is left with physical fields only and we can define a consistent notion of local redefinitions of these physical fields.
This will prove necessary in order to properly identify symplectic deformations that are physically equivalent.
In fact, the quotient in \eqref{normalizer over stabilizer} corresponds to a set of redefinitions that can mix electric and magnetic fields.
Hence they may modify couplings in a way that is irrelevant at the classical level, but that can become physically meaningful when considering quantum corrections.
Therefore, we will use two definitions of equivalence for symplectic deformations: equivalence of the Lagrangians by local field redefinitions exclusively, which will require us to modify the left quotient of \eqref{normalizer over stabilizer}, or equivalence at the level of the equations of motion and Bianchi identities only, giving rise to an $\mathfrak S$-space and a reduced $\mathfrak S$-space, respectively.

Recall that two ungauged Lagrangians of maximal $D=4$ supergravity are related by $\mathrm{Sp(56,{\mathbb{R}})}$ transformations $S_M{}^N$ acting on the $\mathrm{E_{7(7)}}/\mathrm{SU(8)}$ coset\footnote{It is worth mentioning that the scalar manifold of maximal $D=4$ supergravity actually is ${{\rm E_{7(7)}}/({\rm SU(8)/\bbZ_2}) }$ \cite{Cremmer:1979up}.
The extra factor $\mathbb{Z}_{2}$ is due to the fact that spinors, as a
consequence of their interaction with gauge fields through bilinear fermionic
terms, transform according to the double cover of the stabilizer of the scalar manifold itself.}
representatives as \cite{CDFVP,GZ,deWit:2005ub,deWit:2007mt}
\begin{equation}
L(\phi )_{M}{}^{\underline{N}}\rightarrow S_{M}{}^{P}L(\phi )_{P}{}^{%
\underline{N}}.  \label{symplectic trf on coset representatives}
\end{equation}%
If we write the kinetic terms for the vector fields as
\begin{equation}
e^{-1}\mathcal{L}_{\mathrm{vector}}=-\frac{\mathrm{i}}{4}\left( \mathcal{N}%
(\phi )_{\Lambda \Sigma }F_{\mu \nu }^{+\Lambda }F^{+\Sigma \mu \nu }-\overline{%
\mathcal{N}}(\phi )_{\Lambda \Sigma }F_{\mu \nu }^{-\Lambda }F^{-\Sigma \mu
\nu }\right) ,
\end{equation}%
the gauge-kinetic function $\mathcal{N}_{\Lambda \Sigma }$ transforms as a consequence of \eqref{symplectic trf on coset representatives} according to:
\begin{equation}
\mathcal{N}\rightarrow (U\mathcal{N}+W)(V+Z\mathcal{N})^{-1},\quad
S_{M}{}^{N}=%
\begin{pmatrix}
U_{\Lambda }{}^{\Sigma } & W_{\Lambda \Sigma } \\
Z^{\Lambda \Sigma } & V^{\Lambda }{}_{\Sigma }%
\end{pmatrix}%
.  \label{transformation of gauge kinetic function}
\end{equation}%
Similar transformation properties hold for moment couplings of field strengths with fermion bilinears.

In the gauged models, the change of symplectic frame also acts on the embedding tensor according to
\begin{equation}
X_{MN}{}^P\to S_M{}^Q\ S_N{}^R X_{QR}{}^S S^{-1}{}_S{}^P.
\end{equation}
This ensures that the $T$-tensor, defined as
\begin{equation}  \label{T tensor definition}
T(\phi)_{\underline M\underline N}{}^{\underline P} =
L^{-1}(\phi)_{\underline M}{}^M L^{-1}(\phi)_{\underline N}{}^N\ X_{MN}{}^P
L(\phi)_P{}^{\underline P},
\end{equation}
and hence the fermion supersymmetry shifts as well as the scalar potential, are independent of the choice of symplectic frame. 
This in turn guarantees that the combination of equations of motion and Bianchi identities is invariant under symplectic transformations.

Now, we have shown that, starting for simplicity with electric gauge generators $t_r$ as in \eqref{electric gauge generators}, any consistent gauge connection $\vartheta_M{}^r$ can be mapped to the standard electric one, $\delta_M{}^r$, by an element $N$ of ${\cal N}_{{\rm Sp(56,\bbR)}}({\rm G_{gauge}})$.
As a consequence, in the original electric frame, which we call \emph{frame 1}, we have two (potentially) inequivalent gaugings giving rise to different $T$-tensors:
\begin{align}
\label{T0 in frame 1}
X^0_{MN}{}^P \equiv \delta_M{}^r\,t_{r\,N}{}^P &\quad \Rightarrow\quad T^0(\phi)_{\underline{MN}}{}^{\underline P}\qquad (frame\ {\it1}),\\[1ex]
\label{Tvth in frame 1}
X^\vartheta_{MN}{}^P \equiv \vartheta_M{}^r\,t_{r\,N}{}^P &\quad \Rightarrow\quad T^\vartheta(\phi)_{\underline{MN}}{}^{\underline P}\qquad (frame\ {\it1}).
\intertext{%
The change of symplectic frame associated with $N^{-1}$ maps $X^\vartheta$ to the electric embedding tensor $X^0$ by definition, but as we just discussed the $T$-tensor is invariant under symplectic transformations and hence \emph{in the $N^{-1}$-frame} the gauging defined by \eqref{Tvth in frame 1} becomes
}
\label{Tvth in frame N-1}
X^0_{MN}{}^P \equiv \delta_M{}^r\,t_{r\,N}{}^P &\quad \Rightarrow\quad T^\vartheta(\phi)_{\underline{MN}}{}^{\underline P}\qquad (frame\ N^{-1}).
\end{align}
As a result the equations of motion and Bianchi identities obtained by \eqref{Tvth in frame 1} are equivalent to those obtained from \eqref{Tvth in frame N-1}.
We stress again that the gauge generators $t_r$ are the same --- and hence, in particular, electric --- in both frames.
We identify the $N^{-1}$-frame as a symplectic frame in which the connection $\vartheta_M{}^r$ is brought back to its standard electric form.

An alternative way of seeing the above discussion, which will prove useful in showing explicitly how to identify the set of truly inequivalent theories, is to start from the electric gauging $X^0$ in \emph{frame 1}, as in \eqref{T0 in frame 1}, and notice that if we apply the $N^{-1}$ transformation \emph{only} to the coset representatives, namely
\begin{equation}  \label{N-1 trf only on coset representatives}
L(\phi)_M{}^{\underline N} \to N^{-1}_M{}^P L(\phi)_P{}^{\underline N},\quad
X^0_{MN}{}^P\ \text{unchanged},
\end{equation}
then the $T$-tensor transforms as (cfr.~\eqref{action of normalizer on X})
\begin{align}  \label{T0 to Tvth} \notag
T^0_{\underline{MN}}{}^{\underline P}\ &\equiv \ L^{-1}{}_{\underline M}{}^M
L^{-1}{}_{\underline N}{}^N\ X^0_{MN}{}^P\ L_P{}^{\underline P} 
\\[1ex]
&\overset{N^{-1}}{\to}\ L^{-1}{}_{\underline M}{}^M L^{-1}{}_{\underline
N}{}^N\ N_M{}^Q N_N{}^R\ X^0_{QR}{}^S\ N^{-1}_S{}^P\ L_P{}^{\underline P}
\notag \\[1ex]
&=\ T^\vartheta_{\underline{MN}}{}^{\underline P}.  
\end{align}
As a result \eqref{N-1 trf only on coset representatives} maps the theory defined by $X^0$ in \emph{frame 1} to the theory defined by $X^0$ in \emph{frame $N^{-1}$}, namely it maps \eqref{T0 in frame 1} to \eqref{Tvth in frame N-1}. 
The gauge kinetic function and moment couplings transform accordingly with the $N^{-1}$ symplectic transformation. 
Clearly the equations of motion and Bianchi identities are not necessarily invariant under  \eqref{N-1 trf only on coset representatives}, as is reflected by the fact that the $T$-tensor changes. 
We then interpret \eqref{N-1 trf only on coset representatives} as a \emph{symplectic deformation}, namely a map between two (potentially) inequivalent gauged models.
The requirement $N\in\mathcal{N}_{\mathrm{Sp(56,{\mathbb{R}})}}(\mathrm{G_{gauge}})$ ensures that $t_r$ are a good choice of gauge generators also after the symplectic deformation, i.e.~they belong to the $\mathfrak{e}_{7(7)}$ algebra of both the old and the new symplectic frame.

\subsection{The quotient space $\mathfrak{S}$}

Now that we have a good general definition of what symplectic deformations are and how they act on fields and couplings in the Lagrangian, we must classify those that yield inequivalent theories.
Depending on the context, what we regard as inequivalent can change.
For instance, for our purposes it is more natural to regard as equivalent those theories that differ from each other only in the value of the gauge coupling constant, even if it is of course a physically relevant quantity.
It is of course straightforward to include it back.
More importantly, as discussed in the previous section we can decide to distinguish between theories that have the same set of equations of motion and Bianchi identities but differ at the quantum level, or regard them as equivalent if we are only interested in the classical regime.
We will begin with the first option, and therefore assume that we have fixed an initial choice of electric frame, so that we can quotient $\mathcal N_{\rm Sp(56,\bbR)}(\rm G_{gauge})$ by the action of local redefinitions of the physical fields only.

There can also be residual `U-duality' symmetries in the gauged models.
However, just like $\rm E_{7(7)}$ dualities (i.e. acting on both scalars and vectors) do not show up in \eqref{inequivalent Lagrangians}, but only $\rm E_{7(7)}$ redefinitions of the scalar fields appear, also here residual $\rm E_{7(7)}$ dualities do not play any role in restricting the space of symplectic deformations.

Let us take two transformations $N,\ N^{\prime }\in \mathcal{N%
}_{\mathrm{Sp(56,{\mathbb{R}})}}(\mathrm{G_{gauge}})$, related by:
\begin{equation}  \label{N and N'}
N = E\, N^{\prime }G,\quad E\in\mathcal{N}_{\mathrm{E_{7(7)}}}(\mathrm{%
G_{gauge}}),\quad G\in\mathcal{S}_{\mathrm{GL(28,{\mathbb{R}})}}(X^0).
\end{equation}
Here we have chosen $G$ to reflect a local redefinition of the vector fields, whose effect on $X^0$ is at most an overall rescaling.
We will now show that this is the right set of $\rm E_{7(7)}$ transformations and local field redefinitions yielding equivalent theories, up to the action of parity to be discussed momentarily.
Substituting in \eqref{T0 to Tvth} we get:
\begin{align}  \label{T tensor transformation with E7 and GL28}
T^0_{\underline{MN}}{}^{\underline P}\ &\overset{N^{-1}}{\to}\ (L^{-1}
E\,N^{\prime }G )_{\underline M}{}^M (L^{-1} E\,N^{\prime }G )_{\underline
N}{}^N\ X^0_{MN}{}^P\ (G^{-1} N^{\prime -1} E^{-1} L )_P{}^{\underline P},
\end{align}
and at the same time the vector kinetic terms and moment couplings transform with $N^{-1}$.
The $\mathrm{E_{7(7)}}$ transformation $E_M{}^N$ can be reabsorbed in the scalar fields, together with a compensating SU(8) transformation acting on fermions, and therefore it does not affect the physics. Since we have required that the action of $G$ on $X^0$ is trivial up to an overall rescaling, so that 
\begin{equation}  \label{action of elements of hat N GL28}
G_{M}{}^Q G_N{}^R\ X^0_{QR}{}^S\ G^{-1}_S{}^{P} \propto X^0_{MN}{}^P,
\end{equation}
we can reabsorb the rescaling in the gauge coupling constant, and similarly $G_M{}^N$ can be reabsorbed in a local field redefinition of the electric vectors $A_\mu^\Lambda$ in the covariant derivatives and in the non-minimal couplings:
\begin{equation}  \label{GL28 redefinition of Amu}
A_{\mu}{}^\Lambda \to A_{\mu}{}^\Sigma G_\Sigma{}^\Lambda.
\end{equation}
We conclude that $N$ and $N^{\prime }$ in \eqref{N and N'} define the same gauged theory up to local field redefinitions and rescalings of the gauge coupling constant. 

When we can choose an electric frame such that $t_r$ does not contain gaugings of the Peccei--Quinn symmetries (corresponding to the $C_r$ blocks in \eqref{electric gauge generators}), then 
$\mathrm{G_{gauge}}\subset\mathrm{GL(28,{\mathbb{R}})}$.
We may then expect to be able to to reabsorb all elements of $\mathcal{N}_{\mathrm{GL(28,{\mathbb{R}})}}(\mathrm{G_{gauge}})$ in local field redefinitions, but this is not necessarily true in general. Focussing for simplicity on the gauge structure constants, there is the possibility that some elements of the centralizer in $\mathrm{GL(28,{\mathbb{R}})}$ commute with $f_{rs}{}^t$, namely their upper-left block $g_r{}^s$ satisfies
\begin{equation}
g_s{}^u f_{ru}{}^v g^{-1}_v{}^t = f_{rs}{}^t,
\end{equation}
but still $g_r{}^s$ is not proportional to the identity matrix and $g_r{}^u f_{us}{}^t \,\,/\hspace{-2.1ex}\propto f_{rs}{}^t$. Since the $T$-tensor does not contain any contraction with vector fields, \eqref{GL28 redefinition of Amu} cannot be used to remove $G_M{}^N$ from \eqref{T tensor transformation with E7 and GL28}, and its effect would be to give a different $T$-tensor than the one defined by $N^{\prime }$. 
A similar argument is valid for the $h_{rs}{}^a$ components of $t_r$ in \eqref{electric gauge generators}.
This means that if semisimple gaugings of maximal supergravity exist, the separate rescalings of the gauge coupling constants for each simple factor would be classified as inequivalent symplectic deformations, unless they can be reabsorbed in $\rm E_{7(7)}$.
No such gaugings are known, but we may take as an example the $\rm SU(2)\times SU(2)$ $\mathcal N=4$ gauged supergravity of \cite{Freedman:1978ra}, where we expect the separate rescaling of the couplings of the two SU(2) factors to be an example of a symplectic deformation in $\mathcal N=4$ gauged supergravity (another example being the de Roo--Wagemans angles \cite{deRoo:1985jh}).
Moreover, for non-semisimple gaugings of maximal supergravity there could also exist $\mathrm{GL(28,{\mathbb{R}})}$ transformations that centralize $\mathrm{G_{gauge}}$ but act on the gauge connection non-diagonally.

Barring a discussion on the $\bbZ_2$ outer automorphism of $\rm E_{7(7)}$ that we postpone to section~\ref{sec:parity}, we arrive at the result anticipated in the introduction, that symplectic deformations are classified by the space
\begin{equation}  \label{S space}
\mathfrak{S }\ \equiv\ \mathcal{S}_{\mathrm{GL(28,{\mathbb{R}})}}(X^0) \
\backslash\ \mathcal{N}_{\mathrm{Sp(56,{\mathbb{R}})}}(\mathrm{G_{gauge}})\
/\ \mathcal{N}_{\bbZ_2\ltimes\mathrm{E_{7(7)}}}(\mathrm{G_{gauge}}),
\end{equation}
where the quotients correspond to local field redefinitions.%
\footnote{$\mathfrak S$ may not yet include all Lagrangians that admit $X^0$ as gauging charges.
In fact, there is also the possibility that two isomorphic algebras $\mathfrak{g}_1,\ \mathfrak{g}_2\subset\mathfrak{e}_{7(7)}$, both admitting consistent gauging, are conjugate in $\mathrm{Sp(56,{\mathbb{R}})}$ but not in $\mathrm{E_{7(7)}}$. 
In such a situation, $\mathfrak{g}_2$ would be mapped to $\mathfrak{g}_1$, and hence possibly its embedding tensor to $X^0$, by a symplectic transformation that does not sit in the normalizer of either algebra.}
Notice that $\mathcal S_{\rm GL(28,\bbR)}(X^0)$, and hence $\mathfrak S$, carry a dependence on the initial choice of electric frame, to the extent that such choice affects the explicit form of $X^0_{MN}{}^P$ (for instance it can affect the Chern--Simons-like couplings $C_r$ in the gauge generators).
Therefore we must specify the explicit form of $X^0$ that we use to compute $\mathfrak S$, or equivalently the specific choice of electric frame in which we construct the gauged theory whose symplectic deformations we want to compute.

\subsection{Non-local field redefinitions and the $\theta$ angles}
\label{sec:non-local redefinitions}

Some symplectic deformations captured by $\mathfrak S$ do not show up in the gauge connection, namely their inverses leave $X^0$ invariant, as a consequence of the fact that in the left quotient we only consider $\rm GL(28,\bbR)$ transformations instead of symplectic ones as in \eqref{normalizer over stabilizer}.
This happens even when $\mathrm{G_{gauge}}$ has maximal dimension, moreover this fact is strictly related to the dependence of $\mathfrak S$ on the choice of electric frame that we pointed out at the end of the previous section.
As anticipated we can choose to quotient away these transformations, reabsorbing them into non-local redefinitions of the vectors, and this can be a good idea especially when treating small gauge groups, where many symplectic transformations in $\mathfrak{S}$ are electric-magnetic dualities of the vector fields that do not enter the gauge connection.
We therefore define the reduced $\mathfrak S$-space
\begin{equation}  \label{reduced S space}
\mathfrak{S}_{\mathrm{red}}\ \equiv\ \mathcal{S}_{\mathrm{Sp(56,{\mathbb{R}})%
}}(X^0)\ \backslash\ \mathcal{N}_{\mathrm{Sp(56,{\mathbb{R}})}}(\mathrm{%
G_{gauge}})\ /\ \mathcal{N}_{\bbZ_2\ltimes{\rm E_{7(7)}}}(\mathrm{%
G_{gauge}}),
\end{equation}
where we also quotient by non-local redefinitions of the vector fields, as long as they stabilize $X^0$ (up to a rescaling of the gauge coupling constant).
$\mathfrak S_{\rm red}$ is completely independent from the choice of symplectic frame, electric or not.
If we construct $\mathfrak S_{\rm red}$ in a frame which is not electric, the left quotient in \eqref{reduced S space} can be reinterpreted as local redefinitions of the larger set of vector fields and two forms that appear for magnetic gaugings.

The definition given in \eqref{reduced S space} is the most direct generalization of the `$\omega$-deformation' of the SO(8) theory, meaning that it contains all and only the deformations of a given gauging that do affect in a non-trivial way the $T$-tensor, and hence the equations of motion and supersymmetry variations.

However, some elements of $\mathfrak{S}/\mathfrak S_{\rm red}$ have a simple and
interesting physical interpretation, and they are precisely those that arise even when $\mathrm{dim}(\mathrm{G_{gauge}})=28$. 
They are associated to the possibility of shifting the $\theta$-term of the action by a field-independent, gauge invariant quantity.
Consider unipotent symplectic matrices $W_{x}$ of the form
\begin{equation}
W_{x}{}_{M}{}^{N}=%
\begin{pmatrix}
\delta _{\Lambda }{}^{\Sigma } & x_{\Lambda \Sigma } \\
& \delta ^{\Lambda }{}_{\Sigma }%
\end{pmatrix}%
,\qquad x_{[\Lambda \Sigma ]}=0.  \label{topological symplectic deformations}
\end{equation}%
These transformations modify the gauge kinetic function $\mathcal{N}(\phi)_{\Lambda \Sigma }$ by a constant shift of its real part, hence shifting the $\theta$-angles by a term proportional to $x_{\Lambda \Sigma }F^{\Lambda }\wedge F^{\Sigma }$. 
A general choice of $W_{x}{}_{M}{}^{N}$ does not stabilize $X^{0}$ because if $x_{\Lambda \Sigma }$ is not gauge invariant, then it also induces a shift in the gauging of Peccei--Quinn symmetries $X_{\Lambda \Sigma \Gamma }^{0}\rightarrow X_{\Lambda \Sigma \Gamma }^{0}-2X_{\Lambda (\Sigma }^{0}{}^{\Delta }x_{\Gamma )\Delta }^{\vphantom{0}}$, which is necessary to compensate for the gauge variation of the shifted $\theta$-angle (In fact, they were called \textquotedblleft Peccei-Quinn symplectic transformation" in \cite{PQ-1}, where their relation with U-duality and the symplectic group has been investigated).
However, if we choose $x_{\Lambda \Sigma }$ to be a gauge invariant matrix, then $W_{x}$ has trivial action on $X^{0}$. We conclude that for any $\mathrm{G_{gauge}}$, unless it is generated by nilpotent matrices exclusively, there is \emph{at least} one symplectic deformation in $\mathfrak S$, associated to the Cartan--Killing form induced by $\mathfrak{e}_{7(7)}$, namely $x_{\Lambda \Sigma }\propto \Theta _{\Lambda }{}^{\alpha }\Theta _{\Sigma}{}^{\beta }\eta _{\alpha \beta }$ in some electric frame.

It is not surprising that such constant $\theta$-angles can be added to the gauged supergravity action, because they clearly do not affect the equations of motion and supersymmetry variations, and they are encoded as symplectic transformations, consistently with the general analysis of \cite{deWit:2005ub,deWit:2007mt}. 
However, what should be stressed is that these $\theta$-terms cannot be reabsorbed in an $\mathrm{E_{7(7)}}$ transformation. Therefore, they parameterize inequivalent (electric) gauged actions in their own right, and we can expect that quantum corrections to the classical actions may in fact depend on the values of these additional $\theta$-angles. As we will see in section~4 a constant, gauge invariant shift in the $\theta$-angle is even possible in the $\mathrm{SO(8)}$ gauged maximal supergravities and it is consistently encoded in $\mathfrak{S}$.

\subsection{Parity\label{sec:parity}}

There is one more identification between symplectic deformations that we must discuss, which is closely related to a parity transformation and whose correct definition for a general gauging is quite subtle.
If a choice $P_M{}^N$ of the ${\mathbb{Z}}_2$ outer automorphism of $\mathrm{%
E_{7(7)}}$ normalizes $\mathrm{G_{gauge}}$, then by defining its action on $N^{-1}_M{}^N$ appropriately we can further quotient by it. 
In fact, $P_M{}^N$ is realized as an anti-symplectic transformation and it is an invariance of the ungauged Lagrangians when combined with a parity transformation \cite{Ferrara:2013zga,TrigianteTOAPPEAR}.
More precisely, we can regard it as encoding the intrinsic parities of the (physical and auxiliary) fields of the theory, and it is therefore crucial to define a parity symmetry.
Up to a local $\mathrm{SU(8)}$ transformation but taking into account a possible $\mathrm{E_{7(7)}}$ shift of the scalar fields, its action on the coset representatives reads:
\begin{equation}  \label{action of P on coset representatives}
P_M{}^N L(\phi)_N{}^{\underline P} = \big(L(\mathcal{P}\phi^{\prime
})_M{}^{\underline P}\big)^*.
\end{equation}
where by $\mathcal{P}\phi$ we denote the action of parity on the 70 spin
0 fields. The complex conjugate arises because underlined indices are in a $%
\mathrm{SU(8)}$ block-diagonal basis, to make contact with the
transformation properties of the fermions.
In the ungauged case, the inverse of $P_M{}^N$ acts on the vector fields together with the explicit action of parity on the Lorentz indices.
The combination of the actions on coset representatives and vector fields leaves the kinetic terms invariant, and this fact generalizes to the whole ungauged theory.

The gauged case is more subtle: we must require that $P_M{}^N$ normalizes $\rm G_{gauge}$, but even in this case its action on $X^0$ can be non-trivial.
If we define
\begin{equation}
X^{(P)}_{MN}{}^P\equiv P^{-1}_M{}^Q P^{-1}_N{}^R\ X^0_{QR}{}^S\ P_S{}^P,
\end{equation}
then $X^{(P)} \neq X^{0}$ in general.
Since the field strengths now contain $X^0$, their transformation property under parity would not be consistent if we acted  with $P^{-1}_M{}^N$ on the vector fields and $X^{(P)} \neq X^{0}$.
However, $X^{(P)}$ still defines a consistent gauging of $\rm G_{gauge}$, therefore there must exist a symplectic matrix $Q_M{}^N $ whose action on $X^0$ is equivalent to that of $P^{-1}_M{}^N$:
\begin{equation}
X^{(P)}_{MN}{}^P = Q_M{}^Q Q_N{}^R\ X^0_{QR}{}^S\ Q^{-1}_S{}^P
,\quad Q_M{}^N\in \mathcal N_{\rm Sp(56,\bbR)}(\rm G_{gauge}).
\end{equation}
Hence, the anti-symplectic transformation
\begin{equation}
\hat P^{-1}{}_M{}^N \equiv Q^{-1}_M{}^P\ P^{-1}{}_P{}^N 
\end{equation}
can act consistently on the vectors and their field strengths.
Notice that this analysis also implies that a parity symmetry is present in the gauged theory only if $Q_M{}^N$ can be reabsorbed in other field redefinitions.

We can now repeat the analysis of equations \eqref{N and N'} and \eqref{T tensor transformation with E7 and GL28}, but taking the relation
$N = P^{-1} N' \hat P^{-1}$ as a starting point, and conclude that the parity identification is
\begin{equation}
N \simeq P N \hat P,
\end{equation}
which is also consistent with the fact that $N$ is symplectic.
Notice that the squares of $P$ and $\hat P$ are trivial up to field redefinitions, more precisely
\begin{align}
 P^2 &\in \mathcal N_{\rm E_{7(7)}}(\rm G_{gauge}),\\
 \hat P^2 &\in \mathcal S_{\rm Sp(56,\bbR)}(X^0).
\end{align}
This is already sufficient for $\mathfrak S_{\rm red}$.
If we only quotient by the set of local redefinitions of the physical fields in an electric frame, in order to obtain an appropriate parity identification we must require that $\hat P$ acts on physical and dual vectors separately.
Notice that, as opposed to $\mathrm{E_{7(7)}}$ redefinitions of the scalar fields, the action of $P$ on $N$ is always combined with $\hat P$ on the other side. 
We prefer anyway to include the parity identification in the right quotients of (\ref{S space},\,\ref{reduced S space}) by a slight abuse of notation (recall that in our notation it is  $N^{-1}$ that belongs to $\mathfrak S$).

The parity identification is guaranteed to exist, for instance, for gauge groups contained in $\mathrm{SL(8,{\mathbb{R}})}$, $\mathrm{SU^\ast(8)}$ and/or $\mathrm{SU(4,4)}$, with $P = \sigma_3\otimes{\mathbbmss{1}}_{28}$ in these symplectic frames.

\section{\label{sec:SO8}The $\mathfrak{S}$ space of \protect\boldmath$\mathrm{SO(8)%
}$}

Let us now consider the explicit example of the $\mathrm{SO(8)}$ gauged maximal supergravity, taken in its standard electric frame with $\rm SL(8,\bbR)$ as electric group. 
We will extend the result of \cite{Dall'Agata:2012bb} on the existence of a family of deformations of this theory, using pure group-theoretical arguments.
Indeed, in this frame $\rm SO(8)\subset GL(28,\bbR)$, and making use of Schur's lemma \eqref{S space} reduces to 
\begin{equation}
\label{S space SO8}
\mathfrak S
 \ =\
 {\cal N}_{{\rm GL(28,\bbR)}}({\rm SO(8)})
 \ \backslash\
 {\cal N}_{{\rm Sp(56,\bbR)}}({\rm SO(8)})\ /\
 {\cal N}_{\bbZ_2\ltimes{\rm E_{7(7)}}}({\rm SO(8)}),
\end{equation}
and no reference to the embedding tensor is in principle necessary.
To simplify the exposition we always write SO(8) when we refer to the gauge group, although what is embedded in $\rm E_{7(7)}$ is actually the centerless group $\rm PSO(8)=SO(8)/\bbZ_2$.
The following analysis will show that $\mathfrak S$ encodes not only the $\omega$-angle of \cite{Dall'Agata:2012bb}, but also the possibility to further deform the (electric) SO(8) gauged supergravity action by a constant, gauge invariant $\theta$-term, as we discussed in section~\ref{sec:non-local redefinitions}. 
This section also gives an explicit example of how the $\mathfrak{S}$ and $\mathfrak S_{\rm red}$ quotients work and yield the correct parameter space of inequivalent $\mathrm{SO(8)}$ gaugings.

The $\mathrm{SO(8)}$ subgroup of $\mathrm{E_{7(7)}}$ that is gauged can be identified, up to $\mathrm{E_{7(7)}}$ conjugation, from the chain of maximal
and symmetric embeddings:
\begin{eqnarray}
\mathrm{E_{7(7)}} &\supset &\mathrm{SL(8,{\mathbb{R}})}\,\,\supset \,\,%
\mathrm{SO(8)}  \label{chain-1} \\
\mathbf{56} &\rightarrow &\mathbf{28}+\mathbf{28}^{\prime }\rightarrow
\mathbf{28}+\mathbf{28}.  \label{chian-1-2}
\end{eqnarray}%
In order to identify the $\mathfrak{S}$ space of symplectic deformations, we
must first of all compute $\mathcal{N}_{\mathrm{Sp(56,{\mathbb{R}})}}(%
\mathrm{SO(8)})$. 
We can start by computing the connected part of the centralizer without the need to resort to any explicit representation, as we now show. 
Later we will use an explicit representation as a cross--check, and to quickly identify any discrete factors.
First of all, the $\mathrm{Sp(56,{\mathbb{R}})}$ adjoint then decomposes as\footnote{Due to a Theorem by Dynkin \cite{Dynkin} and to one of its exceptions ({cfr.~e.g.} Table VII of \cite{Lorente-Gruber}), the embedding of $\mathrm{E_{7(7)}}$ in $\mathrm{Sp(56,{\mathbb{R}})}$ is {maximal} and non-symmetric, and in physics it is known as a remarkable example of the so-called Gaillard--Zumino embedding \cite{GZ}.}\cite{Fortunato:2013cta}
\begin{eqnarray}
\mathrm{Sp(56,{\mathbb{R}})} &\supset &\mathrm{E_{7(7)}} \\
\mathbf{56} &\rightarrow &\mathbf{56}, \\
\mathbf{1596} &\rightarrow &\mathbf{133}+\mathbf{1463}.
\end{eqnarray}%
The adjoint representation of $\mathrm{E_{7(7)}}$ does not contain any $\mathrm{SO(8)}$ singlet, as its adjoint decomposes as $\mathbf{133} \rightarrow \mathbf{28}+\mathbf{35}_{\mathrm{v}}+\mathbf{35}_{\mathrm{s}}+\mathbf{35}_{\mathrm{c}}$ (in our conventions, the adjoint of $\mathrm{SU(8)}$ decomposes as $\mathbf{63}\rightarrow \mathbf{28}+\mathbf{35}_{\mathrm{v}}$). 
Under the chain of embeddings (\ref{chain-1}), the $\mathbf{1463}$ irrep.~of $\mathrm{E_{7(7)}}$ decomposes in the following triality-invariant way:
\begin{eqnarray}
\mathbf{1463} &\rightarrow &\mathbf{1}_{I}+\mathbf{70}+\mathbf{336}+\mathbf{%
336}^{\prime }+\mathbf{720}  \label{row-1} \\
&\rightarrow &\mathbf{1}_{I}+\mathbf{1}_{II}+\mathbf{1}_{III}+2\cdot \left(
\mathbf{35}_{\mathrm{v}}+\mathbf{35}_{\mathrm{s}}+\mathbf{35}_{\mathrm{c}%
}\right) +2\cdot \mathbf{300}+\mathbf{350},  \label{row-2}
\end{eqnarray}%
because it holds that%
\begin{eqnarray}
\mathrm{SL(8,{\mathbb{R}})} &\supset &\mathrm{SO(8)} \\
\mathbf{70} &\rightarrow &\mathbf{35}_{\mathrm{v}}+\mathbf{35}_{\mathrm{c}},
\\
\mathbf{336} &\rightarrow &\mathbf{1}+\mathbf{35}_{\mathrm{s}}+\mathbf{300},
\label{row-3} \\
\mathbf{720} &\rightarrow &\mathbf{35}_{\mathrm{v}}+\mathbf{35}_{\mathrm{c}}+%
\mathbf{300}+\mathbf{350}.  \label{row-4}
\end{eqnarray}%
We observe that (\ref{row-3}) implies that one of the singlets, say $\mathbf{ 1}_{I}$ for definiteness, is in fact a singlet under the whole $\mathrm{SL(8, {\mathbb{R}})}$. 
Moreover, it should be stressed that {three} $\mathrm{SO(8)}$ singlets in the decomposition of the generators of $\mathrm{Sp(56,{\mathbb{R}})}$ actually exist. Thus, a priori and before taking into account any equivalences, a three-parameter family of $\mathrm{SO(8)}$ gaugings of $D=4$ maximal supergravity exists.

Repeating the above analysis including $\mathrm{GL(28,{\mathbb{R}})}$ in the chain of embeddings, namely considering $\mathrm{Sp(56,{\mathbb{R}})}\supset \mathrm{GL(28,{\mathbb{R}})}\supset \mathrm{SL(8,{\mathbb{R}})}\supset \mathrm{SO(8)}$, shows that $\mathbf{1}_{I}$ is actually the $\mathrm{SL(28,{\mathbb{R}})}$ singlet and therefore it generates a $\mathrm{GL(1,{\mathbb{R}})}$. 
Now, the coset $\mathrm{Sp(56,{\mathbb{R}})}/\mathrm{E_{7(7)}}$ has signature $(c,\,nc)=(721,\,742)$ (where \textquotedblleft $c$ \textquotedblright\ and \textquotedblleft $nc$\textquotedblright\ stand for \textit{compact} and \textit{non-compact}, throughout). 
In particular, the $721$ compact generators all belong to the sub-coset $\mathrm{U(28)}/\mathrm{SU(8)}$, where $\mathrm{U(28)}$ and $\mathrm{SU(8)}$ respectively are the maximal compact subgroups of $\mathrm{Sp(56,{\mathbb{R}})}$ and of $\mathrm{E_{7(7)}}$. 
These $721$ generators sit in the $\mathbf{720}+\mathbf{1}$ of $\mathrm{SU(8)}$, which thus, by virtue of (\ref{row-4}), branches as
\begin{eqnarray}
\mathrm{SU(8)} &\supset &\mathrm{SO(8)} \\
\mathrm{U(28)}/\mathrm{SU(8)}:\quad \mathbf{720}+\mathbf{1} &\rightarrow &%
\mathbf{35}_{\mathrm{s}}+\mathbf{35}_{\mathrm{c}}+\mathbf{300}+\mathbf{350}+%
\mathbf{1}.  \label{S-c-2}
\end{eqnarray}%
Thus, among the two remaining $\mathrm{SO(8)}$-singlets $\mathbf{1}_{II}$ and $\mathbf{1}_{III}$, only one suitable linear combination is compact. 
At this point one can easily realize that the $3$-dimensional group manifold parameterized by the three singlets, with signature $\left( c,nc\right) =(1,2)$, is nothing but $\mathrm{SL(2,{\mathbb{R}})}$. 
In fact, we can recognize $\mathrm{SL(2,{\mathbb{R}})}\times \mathrm{SO(8)}\subset \mathrm{Sp(56,{\mathbb{R}})}$ as descending from the chain of two maximal (non-symmetric) embeddings\footnote{The embedding (\ref{emb-1}) is a consequence of a Theorem by Dynkin for \textit{non-simple} maximal $S$-subalgebras \cite{Dynkin}; it is treated \textit{e.g.} in Sec.~10 of \cite{Lorente-Gruber} (see case II a therein).
Suitable non-compact, real forms of such embedding pertain to the Gaillard-Zumino embedding \cite{GZ} in $\mathcal{N}=2$ supergravity coupled to $27$ vector multiplets ($\rm SL(2,\mathbb{R})\times SO(2,26)\subset Sp(56,\mathbb{R})$) and to $\mathcal{N}=4$ supergravity coupled to $22$ matter multiplets ($\rm SL(2,\mathbb{R})\times SO(6,22)\subset Sp(56,\mathbb{R})$).
On the other hand, the embedding (\ref{emb-2}) is a consequence of the same Theorem, but for \textit{simple} maximal $S$-subalgebras ({cfr. e.g.} Sec.~9 of \cite{Lorente-Gruber}), in the case in which the adjoint vector space $\mathbf{28}$ of SO(8) and its Cartan-Killing symmetric invariant form are considered.}
\begin{align}
\mathrm{Sp(56,{\mathbb{R}})}& \supset \mathrm{SL(2,{\mathbb{R}})}\times
\mathrm{SO(28,{\mathbb{R}})},  \label{emb-1} \\
\mathrm{SO(28,{\mathbb{R}})}& \supset \mathrm{SO(8)},  \label{emb-2}
\end{align}%
where the fundamental of $\mathrm{SO(28,{\mathbb{R}})}$ becomes the adjoint of $\mathrm{SO(8)}$.

We must now take into account further discrete factors in the centralizers,
if any exist. 
Computing the discrete factors of the centralizers and normalizers in a representation-independent fashion requires a quite more sophisticated analysis, hence we prefer to switch to the explicit embedding of SO(8) in the $\bf 56$ representation of $\rm Sp(56,\bbR)$, which is given by 
\begin{equation}
{\mathfrak {so}(8)}\ \ni\  t_r =
\begin{pmatrix} 
\Lambda_r & \\
 & \Lambda_r
\end{pmatrix}
\end{equation}
where $\Lambda_r$ are the SO(8) generators in the adjoint representation.
Schur's lemma then implies that any symplectic matrix centralizing SO(8) must be decomposable as the tensor product of a $2\times2$ matrix with $\bbone_{28}$.
This provides a cross-check that the connected part of centralizer of SO(8) in $\rm Sp(56,\bbR)$ is indeed $\rm SL(2,\bbR)$ and proves that there are in fact no further discrete factors.

Since the quotient of the normalizer by the centralizer is contained in the automorphism group of $\mathrm{SO(8)}$, and the latter is clearly contained in $\rm GL(28,\bbR)$, we conclude that
\begin{align}
\mathcal{N}_{\mathrm{GL(28,{\mathbb{R}})}}(\mathrm{SO(8)}) &\simeq \mathrm{%
GL(1,{\mathbb{R}})}\times \mathrm{S}_3, \\
\mathcal{N}_{\mathrm{Sp(56,{\mathbb{R}})}}(\mathrm{SO(8)}) &\simeq \mathrm{%
SL(2,{\mathbb{R}})} \times \mathrm{S}_3,
\end{align}
where we understand that we are quotienting by $\mathrm{SO(8)}$ itself. The discrete S$_3$ is the triality outer automorphism group of $\mathrm{SO(8)}$. 
The above result holds because we can find real matrices representing all the elements of S$_3$: this is accomplished embedding S$_3$ in $\rm GL(28,\bbR)\subset\mathrm{Sp(56,{\mathbb{R}})}$ in terms of the matrices
\begin{equation}
S_{ab} = {\mathbbmss{1}}_2 \otimes \Gamma_{ab}, \quad a,b = \mathrm{v, c, s,}
\end{equation}
where $\Gamma_{ab}$ realize, in the adjoint representation of $\rm SO(8)/\bbZ_2$, the $\mathrm{S_3}$ element exchanging the $a$ and $b$ labels. 
Their explicit form can e.g.~be given in terms of chiral, real $\Gamma^{(2)}$ matrices constructed from a Cliff(8) algebra.

We can parametrize the ${\rm SL(2,\bbR)}$ in ${\cal N}_{{\rm Sp(56,\bbR)}}$ as follows:
\begin{alignat}{2}
\label{G lambda}
G_\lambda &\equiv \begin{pmatrix}
 \lambda & \\ & \lambda^{-1}
 \end{pmatrix}\otimes \bbone_{28},& \lambda \in \bbR \setminus \{0\},\\[1ex]
\label{W theta}
W_\theta &\equiv \begin{pmatrix} 1 & -g^2\theta/{2\pi} \\ & 1\end{pmatrix}\otimes \bbone_{28},& \theta\in\bbR,\\[1ex]
\label{U omega}
U_\omega &\equiv \begin{pmatrix} \cos\omega & -\sin\omega\\\sin\omega & \cos\omega\end{pmatrix} \otimes \bbone_{28},\quad& \omega\in[0,2\pi].
\end{alignat}

Finally, we must compute $\mathcal N_{\rm E_{7(7)}}(\rm SO(8))$.
Direct computation using the explicit expression of the quartic $\rm E_{7(7)}$ invariant $d_{MNPQ}$ \cite{Cremmer:1979up} shows that only the $\bbZ_4$ subgroup generated by $U_{\pi/2}$ of $\rm SL(2,\bbR)$ is contained in $\rm E_{7(7)}$.
Then, we can compute the normalizer by using the fact that the quotient of the normalizer by the centralizer is isomorphic to a subgroup of the automorphism group of SO(8).
We are only interested in the outer automorphisms, but now we notice that the only triality transformation that is allowed is the one exchanging the two spinor representations, because $\mathrm{E_{7(7)}}$ decomposes as $\mathbf{28}+\mathbf{35}_{\mathrm{v}}+\mathbf{35}_{\mathrm{s}}+\mathbf{35}_{\mathrm{c}}$ under $\mathrm{SO(8)}$, but the $\mathbf{35}_{\mathrm{v}}$ are the compact generators of $\mathrm{SU(8)}/\mathrm{SO(8)}$ and therefore their label must stay fixed.\footnote{The transformations $S_{ab}\in \mathrm{GL(28,{\mathbb{R}})}$ can instead exchange any two of the labels v, s, c, because they do not map $\mathrm{E_{7(7)}}$ into itself, but rather they act separately on $\mathrm{U(28)}$ and $\mathrm{Sp(56,{\mathbb{R}})}/\mathrm{U(28)}$, both of which have triality-invariant decompositions.}
Now, the square of an element of the normalizer is necessarily an element of the centralizer, and it is straightforward to see that the only such transformation that belongs to $\rm E_{7(7)}$ is $T\equiv U_{\pm \pi/4}S_{\rm sc}$. 
The explicit expression for $T$ was given in equations (4.16-17) of \cite{DallAgata:2011aa}, in terms of real chiral $\Gamma ^{(2)}$ matrices mapping $\mathbf{8}_{\mathrm{s}}$ indices to $\mathbf{8}_{\mathrm{c}}$ indices and vice-versa, and satisfying appropriate self-duality requirements.
We conclude that
\begin{equation}
\mathcal{N}_{\bbZ_2\ltimes\mathrm{E_{7(7)}}}\simeq \mathrm{D}_{8},
\label{SO8 normalizer in Eseven}
\end{equation}%
where we have already included the outer automorphism of $\mathrm{E_{7(7)}}$.
The dihedral group of order 16 is embedded in the fundamental representation of $\mathrm{E_{7(7)}}$ (in the standard $\mathrm{SL(8,{\mathbb{R}})}$ frame) in terms of its generators:
\begin{equation}
P=\sigma _{3}\otimes {\mathbbmss{1}}_{28},\qquad T=U_{\pi /4}S_{\mathrm{sc}}.
\label{SO8 normalizer trfs in E7}
\end{equation}%
$P$ is antisymplectic, namely $P\Omega P=-\Omega $, but it preserves $d_{MNPQ}$.

At this point we obtain the parameter space of symplectic deformations of $%
\mathrm{SO(8)}$:
\begin{equation}
\mathfrak{S }= \mathrm{GL(1,{\mathbb{R}})}\times \mathrm{S}_3\ \backslash\
\mathrm{SL(2,{\mathbb{R}})}\times \mathrm{S}_3\ /\ \mathrm{D_8,}
\end{equation}
where the reflection element of $\mathrm{D_8}$ acts as the parity identification discussed in section \ref{sec:parity}.

Let us now make contact with the embedding tensor formalism. 
The consistency constraints on the embedding tensor require that it is a singlet under $\mathrm{SO(8)}$. In fact, the $\mathbf{912}$ $\mathrm{E_{7(7)}}$ representation in which $\Theta_M{}^\alpha$ sits contains two $\mathrm{SO(8)} $-singlets in its manifestly triality-invariant decomposition:
\begin{eqnarray}
\mathbf{912} &\to&\mathbf{36}+\mathbf{36}^{\prime }+\mathbf{420}+\mathbf{420}%
^{\prime }  \label{roww-1} \\
&\to&\mathbf{1}_{\theta }+\mathbf{1}_{\xi } +2\cdot (\mathbf{35}_{\mathrm{v}%
}+\boldsymbol{35}_{\mathrm{s}}+\mathbf{35}_{\mathrm{c}}+\mathbf{350}),
\label{roww-2}
\end{eqnarray}
due to the decompositions
\begin{eqnarray}
\mathrm{SL(8,{\mathbb{R}})} &\supset &\mathrm{SO(8)}, \\
\mathbf{36}^{\prime }&\to&\mathbf{1}_{\theta }+\mathbf{35}_{\mathrm{s}},
\label{dec-1} \\
\mathbf{36} &\to&\mathbf{1}_{\xi }+\mathbf{35}_{\mathrm{s}},  \label{dec-2}
\\
\mathbf{420^{\prime }} &\to&\boldsymbol{35}_{\mathrm{v}}+\mathbf{35}_{%
\mathrm{c}}+\mathbf{350}, \\
\mathbf{420} &\to&\boldsymbol{35}_{\mathrm{v}}+\mathbf{35}_{\mathrm{c}}+%
\mathbf{350}.
\end{eqnarray}
The subscripts ``$\theta\,$'' and ``$\xi\, $'' denote the relation to the symmetric tensors $\theta _{AB}$ and $\xi ^{AB}$ that (when positive-definite) define the $\mathrm{SO(8)}$ generators inside $\mathrm{SL(8,{\mathbb{R}})}$, and that we will always assume to be in the standard form $\theta_{AB}\propto\xi^{AB}\propto\delta_{AB}$. 
The original $\mathrm{SO(8)}$ gauged maximal supergravity corresponds to $\theta_{AB}\propto \delta_{AB},\ \xi=0$ and it is electrically gauged in the $\mathrm{SL(8,{\mathbb{R}})}$ frame. 
What we call $X^0$ corresponds to this particular embedding tensor. 
The so-called `$\omega$-deformed' $\mathrm{SO(8)}$ gaugings are then defined by turning on $\xi\neq0$ and they are no longer electric in the $\mathrm{SL(8,{\mathbb{R}})}$ frame. 
This is clearly achieved in the above parametrization by acting on $X^0$ with the matrix $U_\omega$, which consistently matches equation~(20) of \cite{Dall'Agata:2012bb}.
Following our analysis in section~3, we prefer to regard the symplectic deformations as leaving $X^0$ unchanged, but acting on the coset representatives, thus yielding the deformed theories in their respective electric frames.

We now discuss how each transformation affects the theory and how the quotients
work in practice. A convenient parametrization of $\mathrm{SL(2,{\mathbb{R}}%
)}\times \mathrm{S}_3\subset\mathrm{Sp(56,{\mathbb{R}})}$ is given by (recall that it is actually $N^{-1}$ that belongs to $\mathfrak S$)
\begin{equation}\label{explicit form of N for SO(8)}
\mathrm{SL(2,{\mathbb{R}})}\times \mathrm{S}_3\ \ni\ N = U_\omega\
W_\theta\ G_\lambda\ S_{ab}\,,
\end{equation}
where $S_{ab}$ commutes with all the other transformations. 
We will include the action of parity below. 
Consistently with the general discussion, $G_\lambda$ and $S_{ab}$ leave $X^0$ invariant up to a rescaling of the gauge coupling constant $g\to \lambda g$, and their effect on the kinetic terms can be reabsorbed in a local redefinition of the vector fields. 
The case of $S_{\mathrm{sc}}$ is particular, because we may also choose to combine it with a shift in $\omega$ by $\pm\pi/4$ and reabsorb it in the scalar fields as a $T$ transformation. 
In any case, these transformations do not give rise to inequivalent theories except for the above rescaling of the gauge coupling constant.

As already stressed, the transformation $U_\omega$ corresponds to the `$\omega$-deformation' of the $\mathrm{SO(8)}$ gauged maximal supergravity \cite{Dall'Agata:2012bb}. 
Since we can always reabsorb the $S_{\mathrm{sc}}$ part of the $T$ transformation in $\mathrm{GL(28,{\mathbb{R}})}$ by a local redefinition of the vector fields (as noted also in \cite{deWit:2013ija}), the effect of $T$ is to quotient $U_\omega$ by shifts of $\pi/4$ in $\omega$.
Of course $T$ also induces an SU(8) transformation of the fermions.

The unipotent transformation $W_\theta$ has no effect on $X^0$, and therefore it does not influence the $T$-tensor. 
However, its effect on the vector kinetic terms is non-trivial: recalling that according to \eqref{N-1 trf only on coset representatives} the coset representatives transform with $N^{-1}$, we have
\begin{equation}
\mathcal{N}(\phi)_{\Lambda\Sigma} \overset{W^{-1}_\theta}{\to }\mathcal{N}%
(\phi)_{\Lambda\Sigma}+ g^2\frac{\theta}{2\pi}\delta_{\Lambda\Sigma}.
\end{equation}
where $\mathcal{N}_{\Lambda\Sigma}$ already includes the effect of $U_\omega$ and corresponds to the electric gauge kinetic function of the `$\omega$ deformed' $\mathrm{SO(8)}$ theories. 
This transformation clearly represents a {constant, $\mathrm{SO(8)}$ invariant} shift in the $\theta$-angle of the gauge theory, hence it has no effect on the (classical) equations of motion and supersymmetry variations. In fact, it is clear that we can always add a term $\propto \delta_{\Lambda\Sigma} F^\Lambda \wedge F^{\Sigma}$ to the gauged $\mathrm{SO(8)}$ electric action, and the analysis above proves that there is no $\mathrm{E_{7(7)}}$ transformation or local field redefinition that can remove it. 
Taking $\omega=0$ for example, $W_\theta$ can be interpreted as a change of symplectic frame in which the electric group is still $\mathrm{SL(8,{\mathbb{R}})}$, but now embedded in a block \emph{triangular} form (the off-diagonal block only appearing for $\mathrm{SL(8,{\mathbb{R}})}/\mathrm{SO(8)}$). 
However, it is simpler to just consider the whole electric Lagrangian in the standard, block-diagonal $\mathrm{SL(8,{\mathbb{R}})}$ frame, with the addition of the above shift in $\theta$-angle (the couplings of vectors to fermion bilinears are not affected). 
The analysis is basically the same when $U_\omega$ is nontrivial, and we conclude that all `$\omega$-deformed' $\mathrm{SO(8)}$ theories also admit a shift in the $\theta$-term.
Such a shift would provide, for instance, a non-vanishing $\theta$-angle to the action evaluated around the maximally symmetric AdS$_4$ vacuum of these models.

Finally, the identification $N \simeq P N \hat P$ has no effect on $G_\lambda$ and $S_{ab}$, but clearly sends $(\omega,\, \theta) \to (-\omega,\,-\theta)$. 
In the $\rm SL(8,\bbR)$ frame we can take $\hat P=P=\sigma_3\otimes\bbone_{28}$, which reflects the fact that the original SO(8) theory admits a parity symmetry.
Moreover, when we take $\omega=\pi/8$ and $\theta=0$, the outer automorphism of $\rm E_{7(7)}$ in the electric frame is $U_{-\pi/8}(\sigma_3\otimes\bbone_{28})U_{\pi/8}$, but $\hat P$ does not change.
Hence, $P\hat P=U_{-\pi/4}$ which can be reabsorbed in field redefinitions of vector and scalar fields.
This means that we can define a parity symmetry for the $\omega=\pi/8,\ \theta=0$ theory, which curiously exchanges also the two spinor representations of SO(8).

If we keep $\theta=0$ we reproduce the known parameter space for the $\omega$-deformation of the $\mathrm{SO(8)}$ theory, namely $S^1/\mathrm{D_8}$, with identifications $\omega \simeq \pm\,\omega+k\pi/4$, $k\in\bbZ$ and fundamental domain $\omega\in[0,\pi/8]$.
This last result is more rigorously obtained using the reduced space
\eqref{reduced S space}, where $W_\theta$ is removed from the beginning.
It is actually worth stressing that this result is independent on the choice of symplectic frame, as we have
\begin{equation}\label{Sred di SO8}
\mathfrak{S}_{\rm red}\ = S^1/\mathrm{D_8},\qquad\text{%
fundamental domain: }\omega\in[0,\pi/8].
\end{equation}
If we include $\theta$, the $\mathfrak S$-space of symplectic deformations of SO(8) gauged maximal supergravity, in its standard electric frame, is a quotient of an hyperboloid: $\mathrm{(dS_2/ \bbZ_8)/\bbZ_2}$, where we separated the action of $P$. 
If we also impose periodicity in $\theta$, the resulting space has the topology of a two-sphere.

A brief comment is necessary as regards how $\mathfrak S$ changes if we consider unconventional electric frames where SO(8) is embedded in $\rm Sp(56,\bbR)$ in a block-triangular form, inducing also a gauging of Peccei-Quinn transformations.
In this situation some local redefinitions of the gauged vectors may not be available, and this can be the case in particular for the triality transformation that, combined with $T$ acting on the scalars, allows to identify $\omega\simeq\omega+\pi/4$.
The range of $\omega$ is therefore larger in these electric frames if we only allow for identifications associated with local field redefinitions, while $\mathfrak S_{\rm red}$ is always the same and given by \eqref{Sred di SO8}.

\bigskip

\section{\label{Various-Gaugings}Gauge groups in \boldmath $\mathrm{SL(8,{\mathbb{R}})}$, $\mathrm{%
SU^\ast(8)}$ and flat gaugings}

The previous analysis has the advantage of being group theoretical and almost completely independent from the embedding tensor formalism. 
In principle, it could be repeated for any other consistent gauging of maximal $D=4$ supergravity. 
However, such a task would be time demanding, and several complications would arise for non-semisimple gauge groups.

Since the class of symplectic deformations that yield differences at the level of the classical equations of motion is given by deformations of the gauge connection, captured by the reduced space $\mathfrak{S}_{\mathrm{red}}$ \eqref{reduced S space}, we shall focus on the classification of this space for known gaugings.
The task of computing \eqref{reduced S space} can be accomplished straightforwardly by first choosing the set of gauge generators $t_r$, and then solving the \emph{linear} set of equations in $\vartheta_M{}^r$ 
\eqref{vth constraints}. 
This is equivalent to identifying the coset space \eqref{normalizer over stabilizer}, with the further advantage that we can choose any convenient symplectic frame to perform the computation. 
We shall then take into account equivalences due to field redefinitions: the computation of (at least) the connected part of the normalizers in $\mathrm{E_{7(7)}}$ can be also reduced to a set of linear equations, using for example an explicit realization of the structure constants of these groups. 
Then, either physical arguments or the use of tensor classifiers can be used to pin down any residual discrete identifications.

In the next sections we will use the techniques developed so far to identify the space of deformations of the gauge connection for all gauge groups contained in the $\mathrm{SL(8,{\mathbb{R}})}$ and $\mathrm{SU^\ast(8)}$ subgroups of $\mathrm{E_{7(7)}}$. Similarly to the $\mathrm{SO(8)}$ gaugings, all other $\mathrm{G_{gauge}} \subset \mathrm{SL(8,{\mathbb{R}})}$ are defined by two matrices $\theta_{AB},\ \xi^{AB}$ in the $\mathbf{36}^{\prime }$ and $\mathbf{36}$ of $\mathrm{SL(8,{\mathbb{R}})}$ \cite{Cordaro:1998tx, deWit:2002vt, DallAgata:2011aa}. 
The embedding tensor reads 
\begin{equation}  \label{emb tensor theta and xi}
\Theta_{AB}{}^C{}_D = \delta^{C}_{[A}\theta^{\vphantom{C}}_{B]D},\qquad
\Theta^{AB}{}^C{}_D = \delta_D^{[A}\xi_{\vphantom{D}}^{B]C},
\end{equation}
and the consistency constraints impose $\theta_{AC}\xi^{CB}\propto\delta_A^B$ or $\theta_{AC}\xi^{CB}=0$. 
The deformations of the gauge connections that we are going to discuss in the next sections can be always interpreted in terms of an $\omega$ parameter `rotating' $\theta$ and $\xi$, as in the $\mathrm{SO(8)}$ case:
\begin{equation}  \label{omega theta and xi}
\theta_{AB}\to\cos\omega\,\theta_{AB},\quad \xi^{AB}\to\sin\omega\,\xi^{AB}.
\end{equation}
Similar expressions for the $\mathrm{SU^\ast(8)}$ case, in terms of tensors in the $\overline{\mathbf{36}}$ and $\mathbf{36}$ irreps, can be defined. 
We will show that the range of the $\omega$ parameter, when it is allowed, can be very different from model to model.

Before embarking ourselves in this task, however, we may ask whether another well-known class of gaugings of maximal $D=4$ supergravity admits such deformations: the Scherk--Schwarz and Cremmer--Scherk--Schwarz gaugings (CSS
for brevity) \cite{Scherk:1979zr, Cremmer:1979uq, Andrianopoli:2002mf}.
The formalism of equations \eqref{vth constraints}, with $t_r$ defined in terms of the four CSS mass parameters as explained in \cite{Andrianopoli:2002mf}, allows to quickly identify the space $\mathfrak{S}_{\mathrm{red}}$ of deformations of the gauge connection. 
Unfortunately, we find that for the CSS models no such deformation exists, as the connection $\vartheta_M{}^r$ is unique up to the obvious overall rescaling, which is itself a modulus of the theory. 
Therefore, the full $\mathfrak{S}$ space of the CSS gaugings consists exclusively of deformations of the $\theta$-angle of the gauged $\mathrm{U(1)}$ vector field and of a large set of symplectic redefinitions of the ungauged ones.

\subsection[${\mathrm{SO}(p,q)}$ gaugings]{\boldmath${\mathrm{SO}(p,q)}$
gaugings}

The $\mathfrak{S}$ space for the non-compact forms of $\mathrm{SO(8)}$ can be derived by analytic continuation of the $\mathrm{SO(8)}$ theories.
Most of the analysis of Section~\ref{sec:SO8} is unchanged, only with the off-diagonal blocks of the matrices in (\ref{G lambda}--\ref{U omega}) being proportional to the Cartan--Killing invariant form $\eta_{\Lambda\Sigma}$ instead of ${\mathbbmss{1}}_{28}$. One subtlety regards the outer automorphisms of ${\mathrm{SO}(p,q)}$: the analytic continuation will generally map the $\Gamma^{(2)}$ matrices used to define the S$_3$ generators to complex matrices. 
In particular, only for $\mathrm{SO(4,4)}$ it is possible to reconstruct a real $\Gamma_{\mathrm{sc}}$ matrix that can be used to define the $T$ transformation, as was already noted in \cite{DallAgata:2011aa}. 
Other outer automorphisms would be quotiented away in any case, therefore this is the only transformation that can affect the final result. 
The explicit construction of $\Gamma_{\mathrm{sc}}$ for $\mathrm{SO(4,4)}$ shows that the resulting $T$ transformation does indeed belong to $\mathrm{E_{7(7)}}$.
We conclude that the $\mathrm{SO(4,4)}$ gauging has the same (reduced) space of symplectic deformations as $\mathrm{SO(8)}$, namely:
\begin{equation}
\mathrm{SO(8)},\ \mathrm{SO(4,4)}:\qquad \mathfrak{S}_{\rm red}\ = S^1/\mathrm{D_8},\qquad\text{%
fundamental domain: }\omega\in[0,\pi/8].
\end{equation}
The full $\mathfrak{S}$ space also contains a gauge invariant shift in the $\theta$-term proportional to $\eta_{\lambda\Sigma}$.

For $p,q\neq4$, the analysis is still very similar to $\mathrm{SO(8)}$, but the $T$ transformation in equation \eqref{SO8 normalizer trfs in E7} must be substituted with the centralizer ${\mathrm{i}}\sigma_2\otimes\eta$. 
This means that now $\omega$ is identified to $\pm\omega+{k} \pi/2$, $k\in \bbZ$ and we obtain the space
\begin{equation}
\mathrm{SO}(p,8-p),\ p\neq0,4:\qquad \mathfrak{S}_{\rm red}\ =  S^1/\mathrm{D_4},\qquad\text{%
fundamental domain: }\omega\in[0,\pi/4].
\end{equation}
Again, a shift in the $\theta$-angle is also possible. 
The absence of the triality identification is also further confirmed by an analysis of the vacua of the $\mathrm{SO(6,2)}\simeq\mathrm{SO^\ast(8)}$ and $\mathrm{SO(7,1)}$ theories carried out in \cite{DallAgata:2011aa,Catino:2013ppa}: both these gaugings admit vacua preserving their maximal compact subgroups only for $\omega=\pi/4$, which therefore cannot be equivalent to $\omega=0$.

\subsection[The $\mathrm{CSO}(p,q,r)$ and $\mathrm{CSO}^\ast(2p,2r)$ gaugings]{The \boldmath$\mathrm{CSO}(p,q,r)$ and $\mathrm{CSO}^\ast(2p,2r)$ gaugings}

A large class of gaugings that descend from ${\mathrm{SO}(p,q)}$ are In\"on\"u--Wigner contractions of ${\mathrm{SO}(p,q)}$ and $\mathrm{SO^\ast(8)}$, defined in the $\mathrm{SL(8,{\mathbb{R}})}$ and $\mathrm{SU^\ast(8)}$ electric frames respectively \cite{Hull:1984yy,Hull:1984vg,Hull:1984qz,Hull:2002cv}. 
Using the techniques described above, it is rather straightforward to calculate that most of these gaugings do not admit deformations of the gauge connection $\vartheta_M{}^r$. 
The only exceptions are the gaugings $\mathrm{ISO} (p,7-p)\simeq\mathrm{CSO}(p,7-p,1)\subset\mathrm{SL(8,{\mathbb{R}})}$ that, as we will now prove, admit a \emph{discrete} deformation corresponding to the `dyonic' gauging of their seven translational symmetries (with respect to the $\mathrm{SL(8,{\mathbb{R}})}$ frame).
That most CSO and CSO$^*$ gaugings have a trivial reduced $\mathfrak S$  space may come as a surprise, since all of them admit two singlets in the decomposition of the embedding tensor representation $\bf912$.
One singlet corresponds to the $\theta_{AB}$ matrix that defines the gauging (or its equivalent in the $\overline{\bf36}$ of ${\rm SU^\ast(8)}$); the second singlet is given by $\xi^{AB}$ such that $\theta_{AC}\xi^{CB}=0$ (and, again, its analogue for ${\rm SU^\ast(8)}$).
Contrary to the ${{\rm SO}(p,q)}$ case, however, turning on $\xi^{AB}$ does not generally correspond to a mere deformation of the gauge connection, because it also introduces new gauge couplings, giving rise to the families of gaugings \cite{DallAgata:2011aa,Catino:2013ppa}
\begin{align}
\label{dyonic gaugings SL8}
[{{\rm SO}(p,q)}\times{\rm SO}(p',q')]\ltimes N^r\ &\subset\ {\rm SL(8,\bbR)},\\
\label{dyonic gaugings SUs8}
[{\rm SO}^\ast(2p)\times{\rm SO}^\ast(2p')]\ltimes N^r\ &\subset\ {\rm SU^\ast(8)}.
\end{align}
\textit{This shows how having more than one gauge singlet in the decomposition of the embedding tensor is a necessary, but not sufficient, condition for having deformations of the gauge connection.}
We will discuss the gaugings (\ref{dyonic gaugings SL8},\,\ref{dyonic gaugings SUs8}) in the next section.
The only case in which turning on $\xi^{AB}$ gives rise to a symplectic deformation is when $\theta_{AB}$ has only one vanishing eigenvalue, so that turning on $\xi^{AB}$ gauges the same seven nilpotent generators that were already gauged by $\theta_{AB}$.
This gives rise to the ${\rm ISO}(p,7-p)$ gaugings.

The above analysis is confirmed by solving explicitly the gauge connection constraints \eqref{vth constraints} for the $\mathrm{CSO}(p,q,r)$ and $\mathrm{CSO}^\ast(2p,2r)$ gaugings: only $\mathrm{ISO}(p,7-p)$ admit more than one solution up to overall rescalings. 
If we introduce a parameter $\omega$ such that $\omega=0$ corresponds to the electric gauge connection in the $\mathrm{SL(8,{\mathbb{R}})}$ frame and $\omega \neq0$ corresponds to gauging the seven nilpotent generators dyonically, then all non-vanishing values of $\omega$ are equivalent up to a $\bbZ_2\ltimes\mathrm{E_{7(7)}}$ transformation: in fact, $\mathrm{ISO}(p,q)$ admits a continuous outer automorphism corresponding to a rescaling of the nilpotent generators. 
This automorphism is realized in $\mathrm{E_{7(7)}}$ as the only non-compact
generator that is a singlet under $\mathrm{SO}(p,7-p)$. 
More explicitly, the Cartan generators of $\mathrm{E_{7(7)}}$ can be chosen as the diagonal elements of $\mathrm{SL(8,{\mathbb{R}})}$, and the relevant generator has the form (taking $\theta_{A8}=0$)
\begin{equation}  \label{SO(p,7-p) singlet}
\begin{pmatrix}
{\mathbbmss{1}}_7 &  \\
& -7%
\end{pmatrix}%
\end{equation}
in the fundamental representation of $\mathrm{SL(8,{\mathbb{R}})}$. 
It is clear that such generator would rescale $\theta_{AB}$ and $\xi^{AB}$ separately. 
Finally, the sign of $\xi^{AB}$ can be changed by a parity transformation, just like in the $\mathrm{SO(8)}$ case. 
Therefore, the only inequivalent choices correspond to $\xi=0$ or $\xi\neq0$ or, in the language of `$\omega$ deformations', to:
\begin{equation}
\mathrm{ISO}(p,7-p):\quad \omega=0\quad\text{or}\quad\omega\neq0\quad ({\rm mod}\ \pi/2).
\end{equation}
A simple observation excludes the possibility that these two choices can be further identified by some discrete transformation: the $\omega=0$ embedding tensors rescale homogeneously under the action of  \eqref{SO(p,7-p) singlet}, but not under any other non-compact generator of $\rm E_{7(7)}$, while turning on $\omega\neq0$ introduces non-homogeneous terms also under the action of \eqref{SO(p,7-p) singlet}.

The physical relevance of the symplectic deformation of these models is clear in the $\mathrm{ISO(7)}$ case. On the one hand, by an argument given in \cite{DallAgata:2011aa}, the $\mathrm{ISO}(p,7-p)$ theories with $\omega=0$ can at most admit Minkowski vacua (although none are known) because of the homogeneous rescaling of the embedding tensor with respect to a non-compact generator of $\rm E_{7(7)}$.
On the other hand, the $\mathrm{ISO(7)}$ theory with $\omega\neq0$ is known to have an AdS vacuum \cite{DallAgata:2011aa}, which is possible precisely because $\omega\neq0$ breaks the homogeneity property of the embedding tensor. Moreover, \cite{Borghese:2012qm} identified another AdS vacuum of an $\mathrm{ISO(7)}$ gauging of maximal supergravity, and we can now state that it also belongs to the `deformed' model.

\subsection{`Dyonic' gaugings}

The gaugings (\ref{dyonic gaugings SL8},\,\ref{dyonic gaugings SUs8}), when defined in the $\mathrm{SL(8,{\mathbb{R}})}$ and $\mathrm{SU^\ast(8)}$ symplectic frames, necessarily involve magnetic vectors for gauging one semisimple factor, as well as a mix of electric and magnetic vectors for the nilpotent generators. 
They are particularly relevant for the study of Minkowski solutions of gauged maximal supergravity, as it has been found that all $\mathrm{G_{gauge}}\subset\mathrm{SU^\ast(8)}$, together with some more groups in $\mathrm{SL(8,{\mathbb{R}})}$, admit such vacua, with fully or partially broken supersymmetry. 
Moreover, the models allowing for Minkowski vacua are connected to the Cremmer--Scherk--Schwarz gaugings by singular limits in their moduli spaces \cite{Catino:2013ppa}.

Repeating the analysis of previous sections, we find that the only gaugings that admit a symplectic deformation of their gauge connection that is not removed by $\mathrm{E_{7(7)}}$ transformations are of the form
\begin{equation}  \label{SO4SO4T16}
\re(\mathrm{SO(4,{\mathbb{C}})}\times\mathrm{SO(4,{\mathbb{C}})})\ltimes
T^{16},
\end{equation}
where we can choose either two $(p,q)$ real forms for the two factors (in which we obtain a subgroup of $\mathrm{SL(8,{\mathbb{R}})}$), or we can choose $(\mathrm{SO^\ast(4)}\times\mathrm{SO^\ast(4)})\ltimes T^{16}\subset\mathrm{SU^\ast(8)}$.
The only deformation of the gauge connection of these models corresponds to the separate rescaling of the couplings of the two $\re(\mathrm{SO(4,{\mathbb{C}})})$ factors (which also gives an electric-magnetic rotation of the vector fields associated with $T^{16}$).
As usual, it can be parameterized in terms of $\omega$ as in equations (\ref{emb tensor theta and xi},\,\ref{omega theta and xi}).

Let us start with the analysis of the range of $\omega$ for $\mathrm{SO(4,{\mathbb{R}})}^2\ltimes T^{16}$. In terms of \eqref{omega theta and xi}, $\omega=0\ (\text{mod}\,\pi/2)$ corresponds to the ungauging of one semisimple factor, and therefore these values must be excluded. 
Hence, any linear identification on $\omega$ must map ${\mathbb{Z}}\pi/2$ to itself, which means that at most we can expect the equivalence relation $\omega\simeq\pm\,\omega+k\pi/2$. 
We can in fact find the appropriate $\mathrm{E_{7(7)}}$ transformations that yield this result: the change of sign is associated as usual to the action of the outer automorphism of $\mathrm{E_{7(7)}}$, while a shift of $\pi/2$ is induced by ${\mathrm{i}} \sigma_2\otimes{\mathbbmss{1}}_{28} \in \mathrm{E_{7(7)}}$ combined with an $\mathrm{SL(8,{\mathbb{R}})}$ transformation mapping $\theta_{AB}$ into $\xi^{AB}$ and vice-versa. 
This transformation clearly exists since $\theta_{AB}$ and $\xi^{AB}$ have the same signature in the current case and it is clearly associated with the ${\mathbb{Z}}_2$ outer automorphism that exchanges the two $\mathrm{SO(4,{\mathbb{R}})}$ factors. 
The same result holds whenever we take the same two real forms in \eqref{SO4SO4T16}, while in all other cases $\theta_{AB}$ and $\xi^{AB}$ have different signatures, so that we lose one identification, therefore we expect the range of the deformation to be $\omega\simeq\pm\,\omega+k\pi$. 
Summarizing, the range of $\omega$ for these gaugings is
\begin{equation}
\re(\mathrm{SO(4,{\mathbb{C}})}\times\mathrm{SO(4,{\mathbb{C}})})\ltimes
T^{16}:\quad
\begin{cases}
\omega\in(0,\pi/4] & \text{same real form}, \\[1ex]
\omega\in(0,\pi/2) & \text{different real forms}.%
\end{cases}%
\end{equation}

The physical relevance of $\omega$ is most clear for $\mathrm{SO^\ast(4)}^2\ltimes T^{16}\subset \mathrm{SU}^\ast(8)$. 
This gauging admits Minkowski vacua with fully broken supersymmetry for any value of $\omega$, and the masses of all fields are completely determined by a mass formula that effectively includes their moduli dependence \cite{Dall'Agata:2012cp}, \cite{Catino:2013ppa}. 
The masses of the gravitini have even multiplicity, therefore we can define three inequivalent mass ratios that determine the different scales of supersymmetry breaking (the overall scale is set by the gauge coupling constant times the Planck mass, multiplied by a modulus). 
It turns out that only two out of three of these mass ratios are governed by the expectation value of some moduli. 
If we define the four independent gravitino masses to be $M_1,\,M_2,\,M_3,\,M_4$, the ratio that is unrelated to any modulus can be taken to be $M_1\, M_2 / M_3\, M_4$. We find that this ratio is governed by the $\omega$ parameter according to
\begin{equation}
\frac{M_1\, M_2}{ M_3\, M_4} = \tan\omega.
\end{equation}
The above discussion on the range of $\omega$ shows that it is exhaustive to consider this ratio to be in the range $(0,\,1]$, as values greater than one can be mapped back to the fundamental domain by a field redefinition that also has the effect of exchanging $M_1,\,M_2$ with $M_3,\,M_4$.
Sending $\omega\to0$ also has a clear physical interpretation: on the one hand, it corresponds to restoring some amount of supersymmetry, and on the other hand it corresponds to a gauge group contraction that yields the model with $\mathrm{CSO}^\ast(4,4)\simeq\mathrm{SO^\ast(4)}\ltimes T^{16}$ gauge symmetry, which indeed admits Minkowski vacua with $\mathcal{N}=4$ supersymmetry \cite{Catino:2013ppa}.

A small puzzle arises when we notice that the algebras of $(\mathrm{SO(4)}\times\mathrm{SO(2,2)})\ltimes T^{16}$ and of $\mathrm{SO^\ast(4)}^2\ltimes T^{16}$ are isomorphic. 
We may then ask if the above discussion also applies to the former gauging, which can be seen as arising from a contraction of $\mathrm{SO(6,2)}_{\omega=\pi/4}$ along its moduli space and indeed it admits non-supersymmetric Minkowski vacua \cite{DallAgata:2011aa},\cite{Catino:2013ppa}. 
The mass spectra coincide too, but the mass ratio $M_1\, M_2 / M_3\, M_4$ in the $(\mathrm{SO(4)}\times\mathrm{SO(2,2)})\ltimes T^{16} $ model is not regulated by its $\omega$ deformation, as the latter in fact breaks the vacuum condition. 
A full analysis of the identifications between these Minkowski models goes beyond the scope of this paper, and we leave it for future work.

\section{\label{Comments}Comments}

With this work, we presented a detailed procedure to determine the space of symplectic deformations of gauged maximal supergravity.
This clarifies a number of pending issues in understanding such theories.
In particular, it is now clear that the deformations are continuous, because they can be interpreted as non-local field redefinitions needed to change the symplectic frame in which we introduce the gauge couplings $X^0$. 
Hence it seems that charge quantization conditions should not affect the deformation parameters, as 
$\mathfrak S$ simply parameterizes a set of Lagrangians compatible with the gauging $X^0$.
Moreover, in $\mathfrak S$ two Lagrangians are regarded as equivalent when they can be mapped to each other by local field redefinitions, hence also these identifications are not affected by the discretization of duality groups. 

It is also interesting to see that the space of inequivalent deformations of the SO(8) theory is not limited to the $\omega$ parameter introduced in \cite{Dall'Agata:2012bb}, but that there is another parameter, related to the introduction of a $\theta$-term in the Lagrangian, which cannot be reabsorbed in $\rm E_{7(7)}$ dualities or local field redefinitions.
Such a term is irrelevant at the classical level, but it can affect quantum corrections and therefore it can be also relevant for the dual field theory beyond the large $N$ approximation.

There are many aspects that still deserve a better study.
The first one is obviously the generalization of the procedure described here to the case of models with $\mathcal N<8$.
We expect that this could be easily done in the case of models that include only the gravity multiplet, adapting the dimensions of the symplectic group, the duality group and the dimension of the group of linear transformations of the vector fields contained in the same multiplet.
On the other hand, we probably need more care and a refined analysis for generic gaugings in models where the gravity multiplet couples to other  matter multiplets.
In fact, in this case, only the field redefinitions involving scalar fields in the gravity and vector multiplets will have a non-trivial effect also on the vector fields.
It is actually straightforward to see that the $\omega$ parameter survives various truncations of maximal supergravity \cite{Tarrio:2013qga,Anabalon:2013eaa,Lu:2014fpa}, but that its range changes also according to the number of supersymmetries preserved and the different matter couplings of the truncated theory. 
We obviously expect that when applied to $\mathcal N=4$ theories, our general procedure include the de Roo--Wagemans angles \cite{deRoo:1985jh}.
Even if details may vary, it is clear that the rule of thumb to identify symplectic deformations is to classify duality redefinitions of the vector fields that are compatible with the chosen embedding of the gauge group in the symplectic group, and then quotient by local field redefinitions (or a larger set of transformations if we are only concerned with the classical theory).

\begin{table}[ht]
	\begin{center}
	\begin{tabular}{||c|c|c||c|c|c||}\hline
		G & \textbf{R} & ${\cal N}$ & G & \textbf{R} & ${\cal N}$ \\\hline\hline
		E$_{7(7)}$ & \textbf{56} & 8 & E$_{7(-25)}$ & \textbf{56} & 2 \\\hline
		SO$^*(12)$ & \textbf{32} & 2,\,6 & SO(6,\,6) & \textbf{32} & 0 \\\hline
		SU(3,\,3) & \textbf{20} & 2 & SL(6,\,${\mathbb R}$) & \textbf{20} & 0 \\\hline
		SU(1,\,5) & \textbf{20} & 5 & Sp(6,\,${\mathbb R}$) & \textbf{14}' & 2 \\\hline
		[SL(2,\,${\mathbb R})]^3$ & (\textbf{2},\,\textbf{2},\,\textbf{2}) & 2 & SL(2,\,${\mathbb R}$) & \textbf{4} & 2\\\hline
	\end{tabular}
	\end{center}
	\caption{Simple, non-degenerate groups of type E$_7$. We list the relevant symplectic representations \textbf{R} of G and the number of supersymmetries of the corresponding supergravity theory. 
	Note that in the STU model G $=$ [SL(2,\,${\mathbb R})]^3$ is \textit{semi-simple}, but its \textit{triality symmetry} \protect\cite{stu,stu2} makes it \textquotedblleft effectively simple\textquotedblright \cite{Marrani:2010de}.}
	\label{label}
\end{table}

An interesting case where the procedure described here could be applied with obvious modifications is the one of supergravities with duality groups of type E$_{7}$ \cite{brown,Duff-FD-1,FMY-FD-1}, whose simple, non-degenerate cases are listed in Table 1 (For the difference between degenerate and non-degenerate cases see \cite{Ferrara:2012qp}).
The existence of a symplectic quadratic form and of a unique quartic invariant satisfying suitable constraints for the representation $\mathbf{R}$ of the vector fields under the action of the duality group suggest that most of the results we found in the maximal theory can be reproduced in these models.
In fact, at least in the case where the vectors sit in an irreducible representation, as a consequence of a theory by Dynkin \cite{Dynkin}, the existence of the symplectic form implies the maximal embedding of the group $\rm G$ of duality symmetries:
\begin{equation}
	{\rm G} \subset \rm{Sp}(\rm{dim}_{\mathbb R}\mathbf{R},{\mathbb R}).
\end{equation}
This subset of theories includes all extended supergravities with symmetric scalar manifolds, where we excluded the quaternionic fields of $\mathcal N=2$ theories.
We may then propose as a definition of $\mathfrak S$ the quotient
\begin{equation}
\mathcal S_{{\rm GL}(n_v,\bbR)}(X^0_{MN}{}^P,\,f_M^{(i)})\ \backslash\ 
\mathcal N_{\rm Sp(\rm{dim}_{\mathbb R}\mathbf{R},{\mathbb R})}(\hat{\rm G}_{\rm gauge})\ /\ 
\mathcal N_{\rm \bbZ_2\ltimes G }(\hat{\rm G}_{\rm gauge}),
\end{equation}
where $n_v$ is the number of vectors, $X^0_{MN}{}^P$ denotes the embedding tensor for the subgroup $\rm \hat G_{gauge}\subset G_{gauge}$ that is embedded in $\rm G$, while $f_M^{(i)}$ are Fayet--Iliopolous terms, the index $i$ being inert under symplectic transformations.
We may as well define $\mathfrak S_{\rm red}$ by substituting ${\rm GL}(n_v)$ with ${\rm Sp(\rm{dim}_{\mathbb R}\mathbf{R},{\mathbb R})}$ in the left quotient.
For instance, this definition correctly reproduces the $\omega$ deformation of the gauged STU model obtained as a truncation of $\rm SO(8)$ gauged maximal supergravity, recently analyzed in \cite{Lu:2014fpa}.
In this case only FI terms are present so that $ \rm \hat G_{gauge}$ is trivial, and the quotients are easily computed.
The fact that the range of $\omega$ is still $[0,\pi/8]$ is also a direct consequence of the above definition, and we can  see that for further `pairwise' truncations the triviality of $\omega$ and the arising of a de Roo--Wagemans angle are clearly encoded in our definition.
Moreover, several new transformations can be identified and they definitely deserve further study.
When hypermultiplets are also considered, one may naively guess that, as far as isometries of the hypermultiplets' scalar manifold are not gauged, the analysis of such theories can be encompassed by the same generalization we expect for the other theories with duality groups of type E$_7$.
Otherwise, we may propose to treat the embedding tensor that gauges isometries of the Quaternionic K\"ahler manifold  similarly to our proposed treatment of the FI terms.
However, more complications can arise, for instance from a careful study of the linear and quadratic constraints on the embedding tensor formalism for generic gauged supergravities, and we leave this interesting issue and related details for further future investigations.

Another point we would like to clarify in the future is the existence of deformation parameters of the gauge connection in the case of gauge groups that do not have dimension 28.
In particular, maximal supergravity imposes severe restrictions on the existence of gauge groups of small dimension (for instance it is impossible to produce a U(1) gauging \cite{deWit:2007mt}) and it would be interesting to see the effect on the structure of their symplectic deformations.

One of the most fascinating aspects of this analysis is the insight we obtained on the possible origin of these deformation parameters, which is still elusive, despite some very interesting attempts \cite{deWit:2013ija,Godazgar:2013pfa,Baron:2014yua}.
We clarified above why we expect it to remain a continuous parameter also beyond the classical regime, at least from the point of view of the four dimensional theory.
In theories like SO(8), $\omega$ cannot be a modulus that has been truncated away in the reduction procedure from some higher dimensional theory. 
Otherwise, this would imply that the performed truncation to four dimensions is not consistent, because the ratios between the cosmological constants of two different vacua of these theories are often $\omega$ dependent.
However, for theories with Minkowski vacua like ${\rm SO}^\ast(4)^2\ltimes T^{16}$, $\omega$ preserves the vacuum condition and the above argument clearly does not hold, so that $\omega$ could be a truncated modulus.
In any case, the relation of this parameter with non-local field redefinitions of the ungauged theory hints to a precise mechanism for its generation and we plan to discuss this in a future publication.

\section*{Acknowledgments}

We thank Bernard de Wit and Mario Trigiante for discussions.
This work was supported in part by the ERC Advanced Grant no. 246974 (\textit{``Supersymmetry: a window to non-perturbative physics''}), by the ERC Advanced Grants no.226455 (SUPERFIELDS), by the MIUR grant RBFR10QS5J, by the FWO - Vlaanderen, Project No. G.0651.11, and by the Interuniversity Attraction Poles Programme initiated by the Belgian Science Policy (P7/37).

\end{document}